\documentstyle[12pt,psfig]{article}
\begin{document}

\vskip 1truecm
\rightline{Preprint PUP-TH-1606}
\rightline{ e-Print Archive: hep-ph/9603384}
\bigskip
\centerline{\Large  Motion of Chern-Simons number at High Temperatures }
\medskip
\centerline{\Large           under a Chemical Potential}
\bigskip
\centerline{\Large Guy D. Moore\footnote{e-mail:
guymoore@puhep1.princeton.edu.  Phone:  609-258-5934 (USA)}
}
\medskip

\centerline{\it Princeton University}
\centerline{\it Joseph Henry Laboratories, PO Box 708}
\centerline{\it Princeton, NJ 08544, USA}

\medskip

\centerline{\bf Abstract}

\smallskip

I investigate the evolution of finite temperature, classical Yang-Mills field
equations under the influence of a chemical potential for Chern Simons
number $N_{CS}$.  The rate of
$N_{CS}$ diffusion, $\Gamma_d$, and the linear response
of $N_{CS}$ to a chemical potential, $\Gamma_\mu$, are both computed; the
relation $\Gamma_d = 2 \Gamma_\mu$ is satisfied numerically and the
results agree with the recent measurement of $\Gamma_d$ by
Ambjorn and Krasnitz.  The response
of $N_{CS}$ under chemical potential remains linear at least to $\mu = 
6 T$, which is impossible if there is a free energy barrier to
the motion of $N_{CS}$.  The possibility that the result depends on 
lattice artefacts via hard thermal loops is investigated by changing the
lattice action and by examining elongated rectangular lattices; provided
that the lattice is fine enough, the result is weakly if at all dependent
on the specifics of the cutoff.  I also compare SU(2) with SU(3)
and find $\Gamma_{\rm SU(3)} \sim 7 (\alpha_s/\alpha_w)^4 \Gamma_{\rm SU(2)}$.



\section{Introduction}

Baryon number is violated in the Standard Model\cite{tHooft}.  While
it is conserved to all orders in perturbation theory, nonperturbative 
effects involving topologically nontrivial gauge and Higgs field 
configurations permit its violation at a rate which, at zero temperature,
is suppressed by an exponent of order $\exp(-4 \pi / \alpha_W)$.  While 
this rate is far too low to have any phenomenologically interesting 
consequences, it should be much larger at high temperatures, perhaps high
enough to explain the matter abundance in the universe. 

I will briefly review baryon number violation in the Standard Model.  Because
of the axial anomaly, the Baryon number current is not conserved, but
satisfies
\begin{equation}
\partial_\mu J_{B}^{\mu} =
\frac{-g^2 N_F}{64 \pi^2} \epsilon^{\mu \nu \alpha \beta}
F_{\mu \nu}^{a} F_{\alpha \beta}^{a} \, ,
\label{baryonnonconserve}
\end{equation}
where $N_F$ is the number of generations of fermions and
$F$ is the SU(2) field strength tensor.  I have left out the 
hypercharge fields, which are irrelevant because they do not have nontrivial
topological properties.  For smooth field configurations the righthand side
of this equation is a total divergence,
\begin{eqnarray}
\frac{g^2}{64 \pi^2} \epsilon^{\mu \nu \alpha \beta}
F_{\mu \nu}^{a} F_{\alpha \beta}^{a} & = & \partial^{\mu} K_{\mu}
\nonumber \\
K^{\mu} & = & \frac{g^2}{32 \pi^2} \epsilon^{\mu \nu \alpha \beta} 
	( F_{\nu \alpha}^{a} A_{\beta}^a - \frac{g}{3} \epsilon_{abc}
	A_{\nu}^a A_{\alpha}^b A_{beta}^c ) \, .
\end{eqnarray}
The charge associated with this quantity, 
$\int d^3x (K_0)$, is called the Chern-Simons number, $N_{CS}$.  It 
possesses the important property that, if the field configuration is pure
gauge at two times $t=t_0$ and $t=t_1$, then 
\begin{equation}
N_{CS}(t_1) - N_{CS}(t_0) = \int_{t_0}^{t_1} dt \int d^3 x \frac{g^2}{64
\pi^2} \epsilon^{\mu \nu \alpha \beta}	F^{a}_{\mu \nu} F^{a}_{\alpha \beta}
\label{deltaNCS}
\end{equation}
is an integer (assuming only that the gauge field is everywhere smooth).
$N_{CS}$ is invariant under small gauge transformations (those which can
be built up infinitesimally) and changes by an integer under large
gauge transformations; the difference between $N_{CS}$ at two times is
gauge invariant, because the righthand side of the last equation is.

The action of a Euclidean process which changes $N_{CS}$ is at least
$\int (1/4) F^{a}_{\mu \nu} F^{\mu \nu}_a \geq 8 \pi^2 / g^2$, and is in
fact even larger, because when $F_{\mu \nu}^a$ is nonzero, the Higgs
field gradient generally isn't either.  If the configuration is of greater
spatial extent than the inverse $W$ mass then the Higgs field gradients
dominate the action, which will be $\gg 8\pi^2/g^2$.  
For configurations of very small spatial extent,  
the asymptotic freedom of the weak coupling
$g$ again makes the action grow.  Therefore, unlike QCD, topology
change in the electroweak theory is exponentially suppressed, as
mentioned above.

However, at finite temperature it is no longer relevant how large the
action of a configuration is; the system is excited and there is energy
available to make topological transitions.  In the broken electroweak
phase there is a free energy barrier to such transitions 
\cite{Manton,McLerran},
which proceed at an exponentially suppressed rate 
determined by the temperature and Higgs mass.
At higher temperatures electroweak symmetry is restored, so the Higgs
gradients no longer inhibit topology changing transitions.  For sufficiently
large configurations, with spatial extent $O(1/g^2 T)$, the free energy
barrier, if any, is parametrically 
order unity, and baryon number may be violated readily
\cite{Rubakov}.  This is relevant because any $B+L$ (Baryon plus Lepton
number) present in the early universe would then be erased, and, during
the electroweak phase transition, any $B+L$ separation by the wall of
a broken phase bubble would produce a baryon number asymmetry, because
the excess on the symmetric phase side would be destroyed, while 
that on the broken phase side would be preserved\cite{CKN,JPT}.

The rate which is relevant to this baryogenesis scenario is probably 
the rate at
which a chemical potential for baryon number, caused by the separation of
baryon number across the wall, induces baryon number changing 
transitions\footnote{only probably, 
because in the realistic problem the chemical
potential may extend over a region of space or time small enough 
that the large volume or large time limit are not attained}.
That is, we need to know
\begin{equation}
\Gamma_{\mu} \equiv  \frac{ T \langle - \partial N_{CS} 
                           / \partial t \rangle _{\mu} }
{ V \mu }  \qquad (\mu \ll T) \, , 
\end{equation}
where by $\langle \rangle_{\mu}$ I mean the expectation value when there
is a chemical potential $\mu$ for $N_{CS}$ (equal to $N_F$ times the
chemical potential for baryon number, by Eq. (\ref{baryonnonconserve})), 
and $V$ is the volume over which the chemical potential exists.

A detailed balance argument shows that this rate is half
the rate for $N_{CS}$ diffusion \cite{KhlebShap},
\begin{equation}
\Gamma_d \equiv \lim_{t \rightarrow \infty} 
\frac{ \langle ( N_{CS}(t) - N_{CS}(0) )^2 \rangle } { V t}
\label{Gammaddefined}
\end{equation}
in the absence of a chemical potential.  The argument is quite simple in
the broken phase; the general case is reviewed in section 2.
The nice thing about $\Gamma_d$ is that it is straightforward to measure it
for the classical bosonic theory at finite temperature by lattice techniques,
and there is reason to believe that the behavior of this classical
system corresponds well to the infrared dynamics of the actual finite
temperature, quantum Yang-Mills Higgs system\cite{Grigorev}.

Based on these ideas, and following previous work\cite{Ambjornetal}, 
Ambjorn and Krasnitz have recently performed extensive
numerical simulations of the lattice Yang-Mills system \cite{AmbKras} 
and have demonstrated
convincingly that there is a lattice spacing independent classical 
value for $\Gamma_d$.  On parametric grounds,
$\Gamma_d = \kappa_d (\alpha_w T)^4$; they find $\kappa_d = 1.09 \pm 0.04$.
This work apparently answers the question relevant to baryogenesis, but
it leaves some open questions.  For instance, because the theory has
linear divergences, the lattice spacing 
independence does not necessarily assure the absence of lattice artefacts.
Also, a physical explanation of what sets the rate is lacking.
It is smaller than the analytic estimate found in
/cite{FarrarShap} by a factor of $40$, and in comparison to the 
length $2l$ of a cubic lattice large enough to eliminate finite volume
corrections in its computation, the rate is about $l^{-4}/100$ (I include
the 2 to account for periodic boundary conditions)\cite{Krasprivate}.  It is 
possible but not certain that this apparent suppression arises from a
(parametrically order unity)
free energy barrier to winding number change, as proposed for instance in
\cite{Philipsen}. 

An alternative technique, which can be used to shed some light on these
matters as well as to check the result of Ambjorn and Krasnitz, is to
apply a chemical potential for $N_{CS}$ on the lattice and look at the
rate at which $N_{CS}$ drifts.  The method encounters some fairly serious 
technical difficulties \cite{Ambjornetal}, which I will discuss at some
length in section 3; I present a resolution, which the numerical
results of section 4 demonstrate to be successful.  I will also show that
there is in fact no free energy barrier to topology change in Yang-Mills
theory at finite temperature.  Section 5 will be devoted to investigations
of possible ultraviolet lattice artefacts in the classical lattice
simultations. 
In section 6 I extend the technique to find the strong $N_{CS}$ diffusion 
rate (I replace the group SU(2) with SU(3)).  In section 7
I conclude.  I present a simple, efficient 
thermalization algorithm which works for both
the SU(2) and SU(3) cases in Appendix A.

\section{The Two Rates}

As mentioned above, there is a general relation between the $N_{CS}$ 
diffusion rate $\Gamma_d$ and the linear response rate to a chemical
potential, $\Gamma_\mu$.  This relation was demonstrated already 
in \cite{KhlebShap}, and although their argument explicitly assumes that 
there is a substantial free energy barrier, or, in the symmetric phase,
that the rate is very small, that assumption is actually immaterial to
their argument, as has been shown more recently in \cite{RubakShap2},
where the relation is derived in full generality.

Here I will present an alternate proof, which displays that the relation
between the two rates only relies on the system's periodicity in $N_{CS}$
and its ergodicity.

Let us assume that the generalized coordinate $N_{CS}$ is in
contact with a large thermal reservoir, which evolves ergodically, so
that for large $t$, $\langle (N_{CS}(2t)-N_{CS}(t))(N_{CS}(t)-N_{CS}(0))
\rangle = O(1)$ does not grow as a positive power of $t$.  
This is the condition that the motion of $N_{CS}$ does not have 
arbitrarily large time correlation and is necessary for a diffusion rate
to be well defined; in fact it immediately follows that 
$\langle (N_{CS}(t) - N_{CS}(0))^2 \rangle = O(t)$.  I also assume
that, for any operator ${\cal O}$, $\langle N_{CS}(t) {\cal O}(0) \rangle 
= O(1)$
does not grow as a positive power of $t$.
(These conditions should generically be satisfied for an ergodic system.) 
In this case, the probability for a system
to go from $N_{CS}(0)$ at time 0 to $N_{CS}(t)$ at time $t$ depends 
at leading order for large $t$ only
on those values and $t$, and the probability that the Chern-Simons
number is $N_{CS}(t)$ at time $t$ is, in terms of the probability
distribution ${\cal P}(N_{CS}(0),0)$ for its value at time 0, 
\begin{equation}
{\cal P}(N_{CS}(t),t) = \int dN_{CS}(0) {\cal P}(N_{CS}(0),0) 
	G_t ( N_{CS} ( 0 ) , N_{CS} ( t ) ) \, ,
\end{equation}
where $G_t ( N_{CS}(0), N_{CS}(t))$ is the probability that the system,
initially at $N_{CS}(0)$, will arrive at time $t$ at $N_{CS}(t)$.
$G_t$ is invariant to a simultaneous shift of both arguments by an integer,
and in the absence of a chemical potential, it is invariant 
on exchange of its arguments (by time
reversal invariance), and, when one argument is 0, is invariant under a 
sign change in the other (by parity invariance).

To determine the rate of response to a chemical potential, I want to 
consider the equilibrium probability distribution when a term $\mu N_{CS}$
is added to the Hamiltonian.  Unfortunately the energy is then
unbounded from below and no equilibrium probability distribution exists.
I will therefore add a term $\epsilon N_{CS}^2$ to the Hamiltonian so
that the theory will be well defined; in fact, in the real theory in
finite volume there is such a term, because, as $N_{CS}$ changes, the
number of fermions slowly builds up and changes the chemical potential.
I take $\epsilon$ to be much less than any other quantity in the problem
and consider the equilibrium 
probability distribution ${\cal P}_{\mu}(N_{CS})$.
The chemical potential changes the Boltzmann factors for different
states, so in relation to the distribution without chemical potential,
\begin{eqnarray}
\frac{ {\cal P}_{\mu}(N_{CS1})}{ {\cal P}_{\mu}( N_{CS2})} & = & 
\frac{ {\cal P}_0 (N_{CS1})}{ {\cal P}_0 (N_{CS2})}
 f_{\rm periodic}(N_{CS1},N_{CS2}) \times  
\exp \frac{-\mu}{T}(N_{CS1} - N_{CS2}) 
\nonumber \\
& \simeq &
\frac{ {\cal P}_0 (N_{CS1})}{ {\cal P}_0 (N_{CS2})} 
( 1 - \frac{\mu (N_{CS1}-N_{CS2})}{T} 
+ \frac{ \mu^2 (N_{CS1}-N_{CS2})^2}{2T^2} ) \times
\nonumber \\
& & \qquad \left( 1 + \frac{\mu^2}{T^2} f_{\rm periodic}^1
(N_{CS1},N_{CS2}) \right) \, ,
\label{pofmu}
\end{eqnarray}
where the function $f_{\rm periodic}$ is periodic on integer changes
of either argument and equals 1 when the arguments coincide.  
Its expansion only contains terms even in $\mu$.

Also, the hopping probability $G_t$ takes on $\mu$ dependent corrections
\begin{equation}
G_{t,\mu} = G_t^0 + \frac{\mu}{T} G_t^1 + \frac{\mu^2}{T^2} G_t^2
\label{expandG}
\end{equation}
where $G_t^1$ satisfies
\begin{equation}
G_t^1 (0,N_{CS}(t))= - G_t^1 (0,-N_{CS}(t))
\label{oddundermu}
\end{equation}
because the chemical potential is odd in $N_{CS}$.
There is also an order $\epsilon$ correction which will not be
important here.

Using the condition that, for the equilibrium distribution (with or
without $\mu$), ${\cal P}(N_{CS})$ is time independent, I find
\begin{eqnarray}
 {\cal P}_{\mu}(0,0) & = & {\cal P}_{\mu} ( 0 ,t)= \int dN_{CS}(0) 
{\cal P}_{\mu}(N_{CS}(0),0) G_{t,\mu} ( N_{CS}(0) , 0 ) \, ,
\nonumber \\
 {\cal P}_{0}(0,0) & = & {\cal P}_{0} ( 0 ,t)= \int dN_{CS}(0) 
{\cal P}_{0}(N_{CS}(0),0) G_{t}^{0} ( N_{CS}(0) , 0 ) \, .
\end{eqnarray}
Dividing through by ${\cal P}(0,0)$, I get the equality
\begin{equation}
\int dN_{CS}(0) \frac{ {\cal P}_{\mu}(N_{CS}(0),0)}{{\cal P}_{\mu} (0,0)} 
G_{t,\mu}( N_{CS}(0) , 0 ) 
= \int dN_{CS}(0) \frac{ {\cal P}_{0}(N_{CS}(0),0)}{{\cal P}_{0} (0,0)} 
G_{t}^0 ( N_{CS}(0) , 0 ) \, ,
\end{equation}
which I can expand in powers of $\mu$, using Eqs. (\ref{pofmu}) and 
(\ref{expandG}).  

At order $\mu$,
\begin{equation}
\int dN_{CS} \frac{{\cal P}_0 (N_{CS},0)}{{\cal P}_0 (0,0)} N_{CS} G^0_t 
(N_{CS} , 0 )
= \int dN_{CS} \frac{{\cal P}_0 (N_{CS},0)}{{\cal P}_0 (0,0)} G^1_t 
(N_{CS} , 0 )  \, ,
\end{equation}
which follows automatically 
because both sides are zero; the first because, in
the theory without chemical potential, $\langle \dot{N}_{CS} \rangle = 0$,
and the second because $G^1_t$ is asymmetric.

At the next order,
\begin{eqnarray}
\int dN_{CS} \frac{{\cal P}_0 (N_{CS},0)}{{\cal P}_0 (0,0)} 
\frac{N_{CS}^2}{2} G^0_t ( N_{CS},0)
& = &  \int dN_{CS} \frac{{\cal P}_0 (N_{CS},0)}{{\cal P}_0 (0,0)} 
N_{CS} G^1_t (N_{CS},0)
  \\
& - & \int dN_{CS} f_{\rm periodic}^1 \frac{{\cal P}_0 (N_{CS},0)}
{{\cal P}_0 (0,0)} G^0_t (N_{CS},0) 
\label{junk1}
\\
& - & \int dN_{CS} \frac{{\cal P}_0 (N_{CS},0)}{{\cal P}_0 (0,0)} 
G^2_t (N_{CS} , 0 ) \, .
\label{junk2}
 \end{eqnarray}
The term on the lefthand side is half the mean squared change in $N_{CS}$
between time 0 and time $t$, in the system without chemical potential,
which is $Vt \Gamma_d/2$; the 
righthand side is $Vt \Gamma_\mu$. 
The first line is the desired relation, $\Gamma_{d}/2 = \Gamma_{\mu}$.
All that must be shown is that the remaining terms do not grow linearly
with large $t$.  

For (\ref{junk1}) this holds because
$f_{\rm periodic}^1  {\cal P}_0 (N_{CS},0) / {\cal P}_0 (0,0)$ is bounded,
and $\int dN_{CS} G^0_t ( N_{CS},0)$ is 1 because it is the probability 
that if the system starts at $N_{CS}=0$, it will end with any value of
$N_{CS}$.  Hence the term as a whole is bounded, independent of $t$, and
cannot grow linearly in the large $t$ limit.

For the second term, (\ref{junk2}), we can see that it is bounded at large
time by considering the evolution of the system with chemical potential,
but when the initial probability distribution is ${\cal P}_0 $, the
equilibrium distribution for zero chemical potential.  Since the distribution
starts out periodic, it must remain periodic, and 
\begin{equation}
{\cal P} (0,t)/{\cal P}(0,0) = \int dN_{CS} \frac{ {\cal P}_0( N_{CS} , 0 )}
{ {\cal P}_0 ( 0 , 0 ) } 
\left( G^0_t (N_{CS},0) + \frac{\mu}{T} G^1_t (N_{CS},0) 
+ \frac{\mu^2}{T^2} G^2_t (N_{CS},0) \right)
\end{equation}
must remain bounded.  The $G^1_t$ term drops because the initial 
distribution is symmetric, and the $G^0_t$ term gives 1; the $G^2_t$
term is the one we want to know, and we see that it is bounded.
In fact, I expect that the probability distribution will 
approach ${\cal P}_0 (N_{CS}) * f_{\rm periodic}$ at large $t$, 
so the two extra
terms (\ref{junk1}) and (\ref{junk2}) probably cancel; but it is
sufficient to show that they remain bounded at large $t$, which I have
now done.  Hence the relation $\Gamma_d = 2 \Gamma_\mu$ follows.

Note that it is essential in the above argument that the typical distance
that $N_{CS}$ moves, over a time long compared to the thermalization time,
is much less than $T/\mu$, a condition on the size of $\mu$.  If there is
a free energy barrier between integer values of $N_{CS}$, then the important
typical distance which $N_{CS}$ moves will be at least 1, and we can only
be assured of linear response to $\mu$ for $\mu < T$.

With a few additional assumptions we can explore the behavior of the 
system for larger $\mu$, when there is a large free energy barrier.  
I will assume that 
the free energy barrier is significantly larger than $T$, and that
the motion of $N_{CS}$ is quite heavily damped, so that if one barrier
is breached, enough of the energy used (and, at nonzero $\mu$, gathered) in
breaching it is lost that the system is unlikely to breach the next barrier
before becoming thermalized in the new local minimum.  In this case,
the probability density will become equilibrated about the local minimum,
and the probability to go over a barrier will be a kinetic prefactor,
times the 
Boltzmann suppression to get to the top of the barrier.  If $\mu$ is a
small fraction of the barrier height, then the shape of the barrier 
will not be modified much, and the kinetic prefactor will be approximately
equal to its equilibrium value.
In this case, the rate of $N_{CS}$ change will be 
\begin{equation}
\langle \dot{N}_{CS} \rangle = V (\Gamma_{\rm forward} 
- \Gamma_{\rm backward})  \simeq 
\frac{\Gamma_d V}{2} \left( e^{ -\mu \over 2T} - e^{\mu \over 2T} \right)
= - \Gamma_d V \sinh \frac{\mu}{2 T}
\end{equation}
which of course has the correct small $\mu$ behavior.  However, as $\mu$ 
becomes on order or greater than $T$, the rate rises, eventually greatly
exceeding the linear extrapolation.  The behavior will depart from
the sinh function when
the energy gained from the chemical potential in hopping a barrier
becomes comparable to the loss from friction, or when the shape of the 
barrier is significantly modified by the chemical potential term.  
Note that if, contrary to
my assumption, the friction is weak, then most barrier jumping events are
followed by another, and the diffusive motion will have
very long time correlations.  This is not observed in the 
simulations.  I will just comment that, in this situation, the very large
$\mu$ behavior of the system gives a rate which differs from the linear 
extrapolation by on order exp(barrier height).

What happens when there is no barrier?  I will mention the simplest
possible model; that $N_{CS}$ moves in a flat potential under a linear
frictional force and a random force to keep it thermalized,
\begin{equation}
\ddot{N}_{CS} = - \gamma \dot{N}_{CS} + f
\end{equation}
where $f$ is independent of $N_{CS}$ and has $\langle f(t) f(0) \rangle = 0$
for $t \neq 0$.  Its magnitude is chosen so that $N_{CS}$ will be 
thermalized at the correct temperature.  Adding a chemical potential
produces a term $-\mu N_{CS}$ on the righthand side.  Averaging over
realizations of $f$, the terms $\ddot{N}_{CS}$ and $f$ drop out of the
determination of $\langle \dot{N}_{CS} \rangle $, 
which is strictly linear in $\mu$.  This behavior will 
eventually break down only because a realistic $f$ does have a correlation
time and a realistic friction is somewhat nonlinear and has some
memory.

Now I will briefly discuss evaluating $\Gamma_\mu$.  It is a linear
response coefficient, and as such we might expect that it can be computed
from a correlator of the equilibrium theory.  In fact this is the case
(for the continuum theory); by considering modifying the Hamiltonian by
$H \rightarrow H + \mu N_{CS}$ and applying the techniques of 
Kubo \cite{Kubo}, we find
\begin{equation}
\Gamma_{\mu} = \lim_{t \rightarrow \infty} \frac{ T 
\langle \{ N_{CS}(t) , N_{CS}(0)\} \rangle} {V} \, ,
\label{Kuboformula}
\end{equation}
where $\{,\}$ are the Poisson brackets,
\begin{equation}
\{ A , B \} = \sum_{i} \frac{ \partial A }{\partial q_i(0)} 
\frac{ \partial B }{\partial p_i(0)}
- \frac{ \partial B}{\partial q_i(0)}
 \frac{ \partial A}{\partial p_i(0)} \, .
\end{equation}

The derivatives 
in Eq. (\ref{Kuboformula}) are with respect to the coordinates and momenta 
at zero time.
The poisson bracket we need then includes expressions such as $\partial
N_{CS}(t) / \partial p(0)$.  To evaluate numerically the contribution to
this expression from one pair of conjugate variables in 
one representative configuration, we would 
have to make a small change to one momentum and then evolve the system 
forward by time $t$, and compare $N_{CS}$ to its value if we had performed
the evolution without changing the momentum.  Since we must sum over all
coordinates and momenta (and on a lattice there are a great many) and over
an ensamble of initial conditions, this will not be a very efficient way of
measuring $\Gamma_{\mu}$.  A more practical way would be to perform all
the terms in the sum on coordinates at once, by adjusting each coordinate
and momentum according to the derivative of $N_{CS}$ with respect to the
corresponding momentum or coordinate.  Provided that we make small changes,
the final $N_{CS}$ should depend linearly on each change, contributions
arising because of the product of two of the changes we made being
suppressed by a power of how small the changes were.  We can also rely
on the ergodicity of the system and apply the adjustments to coordinates and
momenta at each time step, rather than at only one.  This appears to be
the only efficient way of evaluating Eq. (\ref{Kuboformula}); it is,
however, equivalent to evaluating the evolution of the system under the
action of the chemical potential.  So we gain nothing by considering
equilibrium correlators, and the best way to proceed numerically is 
to evolve the system with a chemical potential term added to the
Hamiltonian.

\section{ The Model }

Now I turn to the problem of how to measure $\Gamma_{\mu}$.
I will only be concerned with finding the high
temperature limit.  Far above the electroweak phase
transition temperature the Higgs boson takes on a substantial plasma
mass and probably does not influence the evolution of very infrared
modes, so it will be sufficient 
to evolve the classical Yang-Mills field equations under the
influence of a chemical potential for $N_{CS}$.  

To prepare for the task I review the method of evolving the Yang-Mills
equations when there is no chemical potential.   
The evolution of the fields is not completely specified by the field
equations because they respect the freedom 
to choose a gauge; specifying a gauge renders the evolution unique. 
The most convenient choice is the gauge which
will make $A_0$ everywhere zero; in terms of the gauge fields $A^i$ and their
conjugate momenta, the electric fields $E^i$, and suppressing group 
indicies, the continuum equations of motion in this gauge are
\begin{eqnarray}
\frac{dA^i}{dt} & = & E^i
\nonumber \\
\frac{dE^i}{dt} & = & - D^{j} F^{ji}
\end{eqnarray}
where $F$ is the field strength tensor, and I have
written the space components in terms of a positive three dimensional
metric.  In addition, the time component of the Yang-Mills equations of
motion,
which arises from minimizing with respect to variations in $A_0$, enforces
a first-class constraint,
\begin{equation}
\label{continuumGausslaw}
D^{i} E^{i} = 0 \, .
\end{equation}
On the constraint surface,
this Gauss constraint commutes with the equations of motion, so it need
only be enforced on the initial conditions.  However, it is important that
this property be preserved numerically when we actually evolve 
the field equations.  

It is also well known how to make
a lattice version of the Yang-Mills field equations without a chemical
potential \cite{Kogut}.  The gauge fields become the link matricies
$U^i$, and the electric fields are adjoint fields which I will take to
lie at the beginning of each link, so the link update rule is
\begin{equation}
\dot{U}^i = E^{i}_{\alpha} i \tau_{\alpha} U^i \ .
\end{equation}
(The left index of $U$ lies at the beginning or basepoint of the link and
the right index lies at the end.  To parallel transport $E_i$ to the end
of its link, we then take $U_i^{\dagger} E_i U_i$.)

The Kogut-Susskind Hamiltonian is
\begin{equation}
H = \sum \frac{ E^i_\alpha E^i_\alpha}{2} + \sum_{ \Box } 1 - \frac{1}{2}
{\rm Tr} U_{\Box}
\end{equation}
where the first sum is over all links and the second is over all 
elementary plaquettes,
and $U_{\Box}$ means the product of the link matricies running around
the plaquette.  The coupling constant is absorbed into the inverse
lattice temperature $\beta_L$ which will be used to thermalize the
system; with these conventions $\beta_L$ corresponds to the continuum
value $\beta_L = 4 / (g^2 aT)$, $a$ the lattice spacing.

The Gauss constraint at each lattice point is
\begin{equation}
\sum_{i} E_i(+) - E_i(-) = 0
\end{equation}
where $E_i(+)$ is the electric field on the link 
running forward out of the site and
$E_i(-)$ is on the link running into the site from behind,
parallel transported to the lattice site.  This linear combination of $E$
fields generates a gauge transformation of the $U$ fields 
at the lattice point, and the
requirement that it vanish can be understood as arising from
our using up our freedom to make time dependent gauge
transformations in choosing the temporal gauge.  A simple leapfrog
algorith for the Kogut-Susskind Hamiltonian, as used in
\cite{Ambjornetal,AmbKras}, identically preserves the Gauss constraint.
Krasnitz has recently developed an algorithm, based on a set of
Langevin equations, for thermalizing this 
system to inverse temperature $\beta_L$ while identically preserving 
the Gauss constraint; for details see \cite{Krasnitz}.

\begin{figure}
\centerline{\psfig{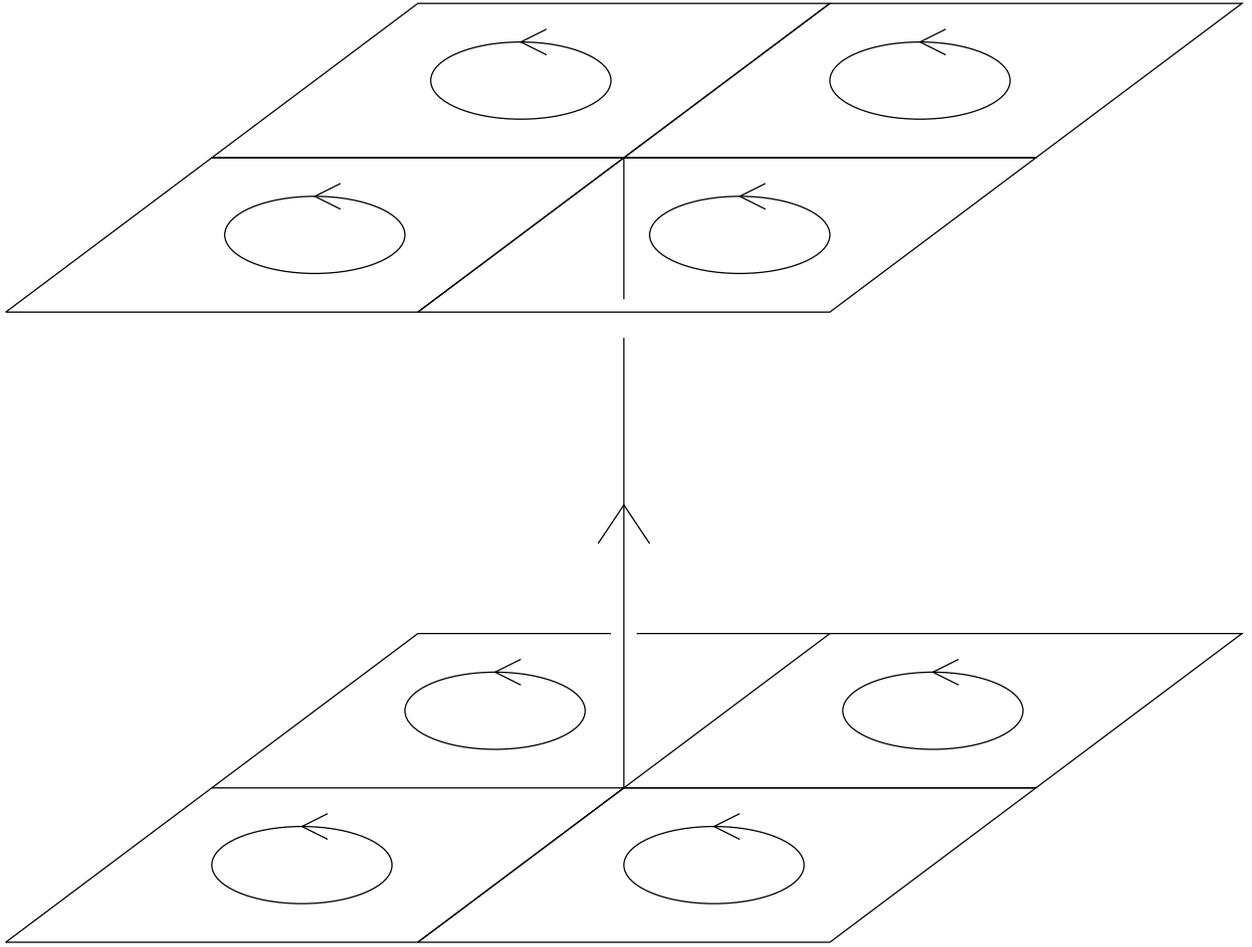}}
\caption{The 8 plaquettes contribute to the magnetic field along the
vertical link.
\label{plaquettes}
}
\end{figure}

To extract the value for $\Gamma_d$ from the realtime evolution, one must
keep track of
the change in $N_{CS}$, which has a gauge invariant continuum definition,
given in Eq. (\ref{deltaNCS}).  In the leapfrog algorithm, where the
$U_i$ are defined at integer multiples of a time stepsize $\Delta t$ and
the electric fields live at half integer values, the simplest lattice 
realization which is invariant under the lattice point group is
\begin{equation}
\label{latticeFFdual}
\frac{\Delta N_{CS}}{\Delta t} = \frac{1}{2 \pi^2} 
\sum_{\rm links} \left( \frac{E_i^{\alpha}(t + \Delta t/2)
+ E_i^{\alpha}(t- \Delta t/2)}{2} \frac{1}{8} \sum_{8 \Box_i} 
\frac{1}{2} {\rm Tr} -i \tau^{\alpha} U_{\Box} \right) \, ,
\end{equation}
where $8 \Box_i$ means a sum over the 8 plaquettes which begin and end at
an endpoint of the link and run in the plane perpendicular to the link,
see figure \ref{plaquettes}.  
(The factors of 2 differ from the continuum version because
the lattice quantities are interms of $\tau$, while continuum fields are
defined in terms of $\tau/2$.)

Now I will discuss the relationship of
this model to the physical, quantum system we hope it will simulate.
The degrees of freedom of the thermal, quantum Yang-Mills system
fall into 3 categories.  There are thermal energy particles with momenta
characterized by the scale $T$, which interact
very weakly with the other excitations, carry almost all of the energy
of the system, and have ultrarelativistic dispersion relations.  There are
more infrared fields with momenta on order the plasma frequency 
$\sim gT$, which
have substantial occupation numbers but interact perturbatively with the
other degrees of freedom, except that forward scattering with the thermal
particles make order unity modifications to their dispersion relations.
And finally there are very infrared modes, with momentum characterized
by the scale $g^2 T$, with large occupation numbers and fully nonperturbative
mutual interactions.

On the lattice there are also three characteristic scales:  excitations
with wavelength order the lattice spacing are weakly coupled, contain almost
all of the system energy, and travel under the lattice dispersion relations;
excitations of wavelength order $a/\beta_L$ interact with each other 
nonperturbatively; and the intermediate scale again has perturbative 
interactions but dispersion relations substantially modified by interactions
with the shortest wavelength modes.
The motion of $N_{CS}$ in the symmetric electroweak phase 
depends on the most infrared modes.  The idea of
Grigoriev and Rubakov is that
in the quantum system these have such
high occupation numbers that they should behave approximately as classical
fields, and should therefore be correctly 
modeled by the classical lattice system\cite{Grigorev}.
If the high energy modes are only important as a thermal bath and for
their corrections to dispersion relations, then the substitution of
classical lattice modes for quantum continuum ones should not
matter much.  

Ambjorn and Krasnitz have also noted that the thermodynamics of the lattice
model bears close resemblance to that of the full quantum system.  In
particular, if one introduces Lagrange multipliers $A_0$ to enforce the
Gauss constraints and then performs the (Gaussian) integration over the 
electric fields, the partition function 
is almost the same as the partition function of the full theory, in
the approximation of dimensional reduction \cite{AmbKras,Dimredpapers}.  
The difference is that the classical, lattice theory has zero {\it bare}
Debye mass.  However, a Debye mass, linearly divergent in the lattice
cutoff, is induced by the high energy excitations.  Its value is 
\cite{Laine}\footnote{In that paper the coefficient is 5 rather than 4,
because they work in Yang-Mills Higgs theory, and the interaction between
$A_0$ and Higgs fields contributes 1 to the coefficient.}
\begin{equation}
m_D^2({\rm lattice}) = \frac{4 g^2 \Sigma T}{4 \pi a} \qquad
\Sigma = 3.1759114 \, .
\end{equation}
The lattice system therefore does in fact reproduce the thermodynamics of
the full theory, for the right value of the lattice constant 
$a$\footnote{This statement is not quite correct, because the lattice 
theory also contains power law divergences in the energy density and in the
expectation values of composite operators; this point is discussed
in \cite{Laine}, which also shows that these divergences do
not interfere with the extraction of physically interesting quantities.}.  
Yang-Mills Higgs theory has $m_D^2 = 5 g^2 T^2/6$, obtained in 
classical lattice Yang-Mills
theory by $\beta_L \simeq 7.8$.  The full standard model has $m_D^2 = 11 g^2
T^2/6$, corresponding to $\beta_L \simeq 17.2$.  The independence of the
rate $\Gamma_d$ on $\beta_L$ therefore
indicates that $\Gamma_d$ does not depend on the Debye mass.
I will return to this point in section 5.  I will also discuss a potential
problem, involving the functional form of the ``hard thermal loops''
induced by the high energy modes.

Now I turn to the problem of adding a chemical potential for $N_{CS}$.
For the continuum equations of motion this is straightforward;
one adds a term $\mu N_{CS}$ to the Hamiltonian, and the
evolution of $A$, $E$ are modified by
\begin{eqnarray}
\frac{dA^i}{dt} & = & \frac{dA^i}{dt}(\mu=0) + \mu \{ A^{i} , N_{CS} \}
	= E^i + 0 \mu
\nonumber \\
\frac{dE^i}{dt} & = & \frac{dE^i}{dt}(\mu = 0 ) + \mu \{ E^i , N_{CS} \}
	= -D^{j} F^{ji} - \frac{ \mu}{16 \pi^2} \epsilon^{ijk} F^{jk}
	= -D^{j} F^{ji} - \frac{ \mu}{8 \pi^2} B^{i}
\label{eqmowithmu}
\end{eqnarray}
where $B^i = (1/2) \epsilon^{ijk} F^{jk}$ is the magnetic field in the $i$
direction.  Fortunately, for smooth fields
\begin{equation}
\epsilon^{ijk} D^{i} F^{jk} = 0
\end{equation}
by the Bianchi identity, and so the additional term still preserves the 
Gauss constraint.  

An equivalent approach is to add to the action a 
term $\theta(t) \epsilon_{\mu \nu \alpha \beta}F^{\mu \nu}_{a} 
F^{\alpha \beta}_{a}$.  If $\theta$ were a constant, then this term would
be a total derivative and would not change the equations of motion; but
if it varies in time, then
when we integrate by parts in time while deriving equations of motion the
time derivative acts on $\theta$, generating the $B^{i}$ term in the 
equations of motion.  Proceeding in this fashion also makes it clear why 
the added term still commutes with the Gauss constraint; we should rederive
the Gauss constraint with this added term in the action, but because
the $0$ index of the antisymmetric tensor is used up by $A_0$, no time
derivatives arise to act on $\theta$, so the Gauss 
constraint is unchanged.

When we try to implement a chemical potential for $N_{CS}$ on
the lattice, we encounter trouble.  The most obvious way is to find
a lattice definition of $N_{CS}$ and add it to the Hamiltonian.  However,
there is no lattice definition of $N_{CS}$ which gives the 
desired behavior under
gauge transformations.  $N_{CS}$ should be invariant under small gauge
transformations and change by an integer under large ones; but on a lattice,
a gauge transformation is a choice of one member of the gauge group at
each lattice site, and since there are finitely many lattice sites and
the gauge group SU(2) is path connected, {\it any} gauge transformation
can be built up infinitesimally.  Therefore, either $N_{CS}$ must be
a constant, or it must vary continuously under gauge transformations.
Also, we would like $N_{CS}$ to change by the same amount on each application
of the same gauge transformation; but any lattice $N_{CS}$ is a function
of finitely many variables, the $U_i$, which live on compact manifolds, and
any continuous function on a compact domain is bounded.  If it changed
by some constant amount under a gauge transformation, then repeated 
application of the gauge transformation would drive it to infinity, a
contradiction.

A second idea for implementing the chemical potential on the lattice
\cite{fromNeil} is to use this definition of the lattice 
$\epsilon^{\mu \nu \alpha \beta} F_{\mu \nu} F_{\alpha \beta} $
and the alternative implementation of the chemical potential in which we
add $\theta(t) \epsilon^{\mu \nu \alpha \beta} F_{\mu \nu} F_{\alpha \beta} $
to the (lattice) action.  The Gauss law, which is derived from the
lattice action, is modified in a way such that it will be preserved by
the new equations of motion.

This technique suffers from a new problem on the lattice, which is that
the lattice definition of $ \epsilon^{\mu \nu \alpha \beta} 
F_{\mu \nu} F_{\alpha \beta} $, Eq. (\ref{latticeFFdual}), is not a total
derivative.  The simplest counterexample configuration contains just
three non-identity link matricies; two, with values $1 + \epsilon i \tau_1$
and $1 + \epsilon i \tau_2$, lie on the two most removed links of
a plaquette perperndicular to some link, and the third,
with value $1 + \epsilon i \tau_3$, lies along the link, backwards one
time step (see figure \ref{counterexample}).  
The space integral of $ \epsilon^{\mu \nu 
\alpha \beta} F_{\mu \nu} F_{\alpha \beta} $ for this configuration is 
nonzero even though all links at large distances are the unit link.  
The definition is therefore not a total derivative\footnote{For any definition
of the space integral of $\epsilon^{\mu \nu \alpha \beta} 
F_{\mu \nu} F_{\alpha \beta}$ which consists of a sum over lattice sites
or links of some local gauge invariant operator, a similar counterexample can
always be found; find three links which all appear in the evaluation of
one, but only one, site or link, and shift their link matricies from the
identity using three orthogonal Lie algebra elements; 
then they will contribute to $\int F \tilde{F}$ at this point but
not at any other, and the definition will not be a total derivative.
Note that the nonabelian nature of SU(2) is essential.}.  This is a serious
problem, because it means that even if $\theta$ is constant, the 
$\theta \epsilon^{\mu \nu \alpha \beta} F_{\mu \nu} F_{\alpha \beta} $
term will change the dynamics of the system; when $\theta$ varies with time, 
then at late times it will become large, and these (spurious) changes
will actually dominate the dynamics.

It is not surprising that this technique did not work.  The argument
in the continuum theory that $\epsilon^{\mu \nu \alpha \beta} 
F_{\mu \nu} F_{\alpha \beta}$ is a total derivative relies on the
fields being smooth, a concept which is lost on the lattice, and its
failure to be a total derivative explains why $N_{CS}$ is not well defined.
Recall that the chemical potential for $N_{CS}$ emerged anyway by integrating
out chiral fermions at nonzero chemical potential, and that it is
impossible to implement chiral fermions on a lattice, a fact which may
be related to our problems here.

Another idea, considered in \cite{Ambjornetal}, is to abandon the hope of
deriving equations of motion from a Hamiltonian and to try to find a lattice
implementation of the continuum equations of motion, Eq. (\ref{eqmowithmu}).
A natural choice for $B_i$ is the average of the eight plaquettes used
in Eq. (\ref{latticeFFdual}); the equations of motion are then
\begin{eqnarray}
U^i(t + \Delta t) & = & \exp(i \Delta t \tau_{\alpha} E^i_{\alpha}) U^i ( t)
\nonumber \\
E^i_{\alpha} (t + \Delta t/2) & = & E^{i}_{\alpha} ( t - \Delta t/2) 
- \sum_{4 \Box_i} \left( \frac{1}{2} 
{\rm Tr} (-i \tau_{\alpha}) U_{\Box} \right)
- \frac{ \mu}{16 \pi^2} \sum_{8 \Box_i} \left( \frac{1}{2} 
{\rm Tr} (-i \tau_{\alpha}) U_{\Box} \right) .
\label{latticeeqmowithmu}
\end{eqnarray}
Here $\sum_{4 \Box_i}$ means a sum over the 4 plaquettes which contain the
link $i$, with orientation so as to contain $U_i$ and not $U^{\dagger}_i$,
and $\sum_{8 \Box_i}$ has the same meaning as previously.  This technique
is nice in that it corresponds to the physical meaning of a chemical
potential term, that the $E$ fields should be modified in accordance with
the $B$ fields so that the energy in the fields, $E \cdot E/2$, is
modified by $E \cdot \delta E = -\mu \Delta t E \cdot B / (2 \pi^2) 
= -\mu \Delta N_{CS}$.  

\begin{figure}
\centerline{\psfig{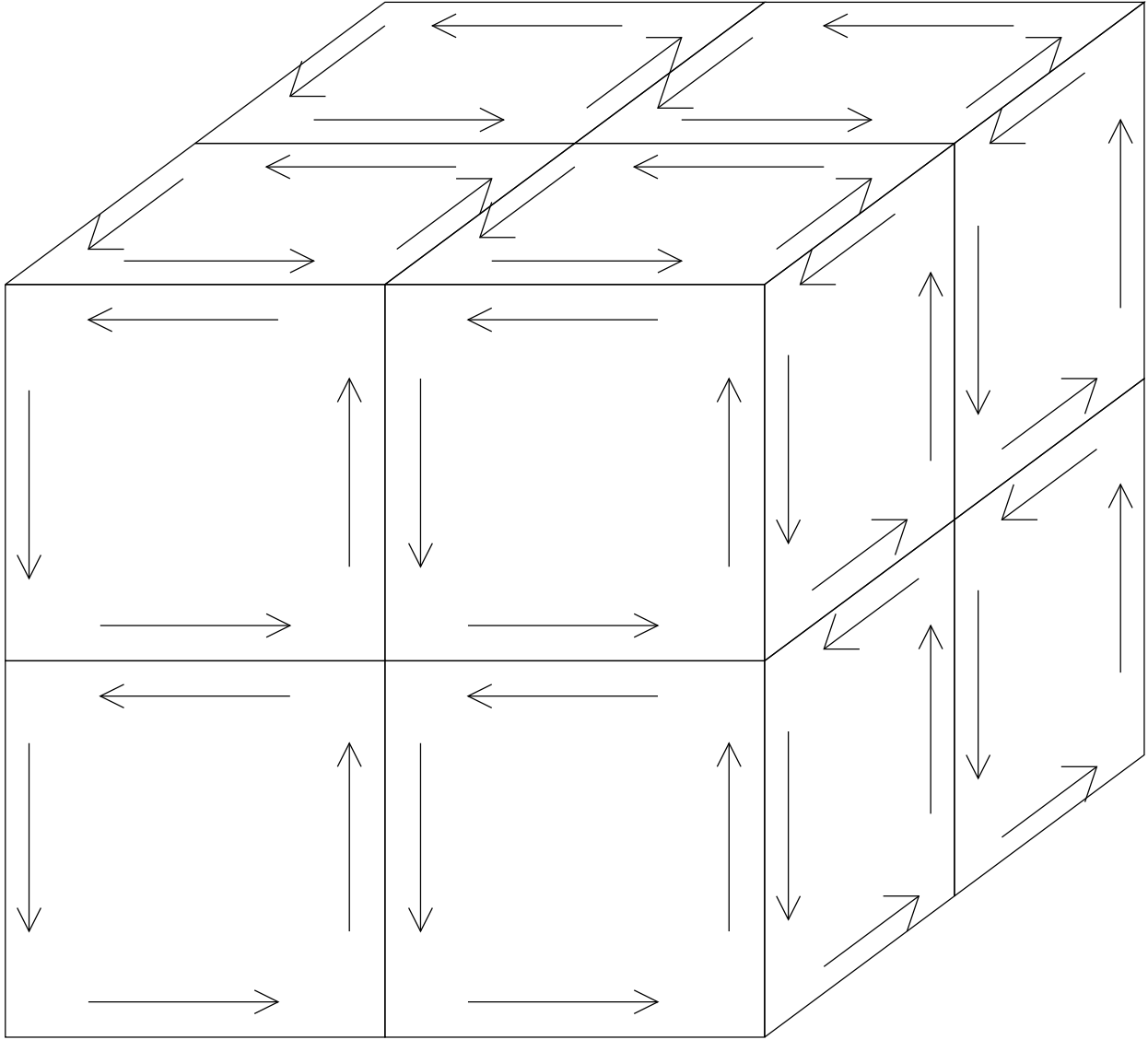}}
\caption{Contributions to $D_i E_i$ at a lattice point arise from 
the surface of the smallest box enclosing that point.  Each
link on the surface is traversed once with each orientation; in the
abelian theory their contributions cancel, but in SU(2) theory, nonzero
commutators arise. \label{boxfig}}
\end{figure}

The problem with
this plan is that the evolution does not preserve the Gauss constraints,
so it excites unphysical modes.  Again, the nonabelian nature of
the theory is essential; in the abelian theory, the lattice definition of
$ \epsilon^{\mu \nu \alpha \beta} 
F_{\mu \nu} F_{\alpha \beta} $ is a total divergence, and the space divergence
of the magnetic field at some point, 
which is the sum of all plaquettes on the surface
of a box one lattice spacing around the point 
(see figure \ref{boxfig}), is the
boundary of a boundary, and vanishes identically,
preserving the Gauss constraint.  But in the nonabelian theory, while
$D_i B_i$ is still the sum of all plaquettes around a cube,
there are commutator terms which spoil the cancellation.  That is, $B^i_a$ is
contaminated with nonrenormalizeable operators such as $a^2 D^l D^l B^i,$ 
and $D^i D^l D^l B^i$ need not vanish, since $[ D^i , D^l]$ does not.
Since the extra terms are nonrenormalizeable, 
they will have vanishing influence
on the infrared physics as the lattice spacing is made smaller; but if
they introduce ultraviolet divergences 
in the measured value of $\Gamma_{\mu}$
then they will spoil the calculation.

If we ignore the violation of the Gauss constraint, or if we (dissipatively) 
remove the Gauss constraint violation occasionally or incompletely,
we risk measuring changes in $N_{CS}$ arising from the excitation
of unphysical modes as well as from genuine topology change.  The
problem is that the combination of electric fields which produce a Gauss
constraint violation has essetially no restoring force.  If the chemical
potential term causes this $E$ field to change, then
$E \cdot B$ will be nonzero for the linear combination of $E$ which make up
the Gauss constraint; if there were a restoring force then $E$ would oscillate 
and the time integral of $E \cdot B$ would vanish, but instead $E$ will rise
and the integral need not vanish.  If the Gauss constraint is dissipatively
enforced occasionally, it is quite easy numerically to establish how much
of any accumulated $N_{CS}$ corresponded to this process.  Because of the
chemical potential, the system energy shifts by $\mu \Delta N_{CS}$.
Some of this energy goes into exciting Gauss constraint
violations; the amount is $\mu$ times the amount of $\Delta N_{CS}$ which
arose from exciting the unphysical modes.  If the Gauss constraint
violations are quenched, then the loss of energy in the quenching 
corresponds to the amount
of spurious $\Delta N_{CS}$.  Numerically I find that this amount is
a substantial share, if the Gauss constraint violation is only quenched
occasionally.  Further, a straightforward argument suggests that the problem
should get worse as the lattice spacing is made smaller.  In lattice
units, in terms of the lattice inverse temperature $\beta_L$, the typical
magnetic field strength is $B \sim \beta_L^{-1/2}$, and since the violation
of the Gauss law depends on the nonabelian nature of the theory the 
size of the induced violation must go as 
$B \times B \sim \beta_L^{-1}$.  To keep $\mu$ fixed in physical
units as we change the lattice spacing, $\mu$ should go as $\beta_L^{-1}$,
so the induced unphysical $E$ field goes as $\mu B \times B \propto 
\beta_L^{-2}$ and the spurious energy introduced per unit lattice 4-volume
goes as $\beta_L^{-4}$.  The rate of $N_{CS}$ violation per lattice 4-volume, 
for $\mu \beta_L$ fixed, due to genuine infrared topology
changing processes should go as $\beta_L^{-4}$, and so the energy shift
from these processes goes as $\beta_L^{-5}$.  If this naive dimensional
argument is right then, if the Gauss constraint is unquenched or poorly
quenched, the contribution to $\Gamma_{\mu}$ due to excitation of the 
unphysical modes
should grow linearly with inverse lattice
spacing, and no fine lattice spacing limit for $\Gamma_{\mu}$ will be found.

The problem can be understood and cured by considering
the implementation of the Hamiltonian equations of motion, 
Eq. (\ref{eqmowithmu}), more carefully.
When we modify the electric fields we should not modify all fields, but
only the linear combinations orthogonal to those constrained to be zero.
That is, the electric fields can be partitioned into two orthogonal subsets,
the constraints, which I call the $E^{c}$, and the fields orthogonal to the
constraints, which I call the $ E^* $.  At fixed $U$ 
the $E^*$ are in correspondence with the momenta of the cannonical
basis of the observable subspace.  Only these $E^*$ are dynamical, the
$E^c$ should be held zero.  Since $U$ is fixed during the update of
$E$ in the leapfrog algorithm, this means we should modify the update
rule in Eq. (\ref{eqmowithmu}) to
\begin{eqnarray}
\frac{\Delta E^c}{\Delta t} & = & \frac{\Delta E^c}{\Delta t}
(\mu = 0 ) + 0 = 0 
\nonumber \\
\frac{\Delta E^* }{\Delta t} & = & \frac{\Delta E^* }{\Delta t}
(\mu = 0 ) + \mu \{ E^* , N_{CS} \}  \, .
\end{eqnarray}
This is exactly what we would conclude if we defined $\Delta N_{CS}$ as
$E^* \cdot B/(2\pi^2)$, which is equivalent to 
$E \cdot B/(2 \pi^2)$ on the constraint surface, since the $E^c$
are zero there.

We can implement this update by finding an orthonormal basis
for the $E^* $,
\begin{equation}
E^*_\alpha = \sum c_{\alpha i} E_i \, , \qquad 
\sum_i c_{\alpha i} c_{\beta i} = \delta_{\alpha \beta}
\end{equation}
and updating them by changing them by
\begin{equation}
E^*_{\alpha} = E^*_{\alpha} - \frac{ \mu \Delta t}{2 \pi^2} 
\sum_i c_{\alpha i} B_i
\end{equation}
or
\begin{equation}
E_i = E_i - \frac{ \mu \Delta t}{2 \pi^2} \sum_{\alpha} c_{\alpha i} 
	\sum_{j} c_{\alpha j} B_j \, ,
\label{rightway}
\end{equation}
which is basis independent and preserves the Gauss constraint by
construction.  Equivalently, given a complete orthonormal basis of Gauss
constraints $E^c$, 
\begin{equation}
E^c = \sum_{i} d_{\alpha i} E_i \, , \qquad
\sum_{i} d_{\alpha i } d_{\beta i} = \delta_{\alpha \beta} \, \qquad
\sum_\alpha c_{\alpha i } c_{\alpha j} + \sum_{\beta} d_{\beta i} d_{\beta j}
= \delta_{ij} \, ,
\end{equation}
the update is
\begin{equation}
\label{rightwayGauss}
E_i = E_i - \frac{\mu \Delta t}{2 \pi^2} \left( B_i - \sum_{\alpha}
d_{\alpha i}  \sum_j d_{\alpha j} B_j \right) \, .
\end{equation}

This technique as described is impractical.  The problem is that (unlike in
the abelian theory) the linear combinations $E^*$ are in general not
well localized and change with each
time step.  Finding them is a problem in the diagonalization of
sparse matricies, but even if it could be performed efficiently, the number
of operations required to implement Eq. (\ref{rightway}) grows with
the square of the number of lattice sites, which makes it numerically
impractical.  We also cannot implement the algorithm by using 
Eq. (\ref{rightwayGauss}) because the natural,
local basis for the Gauss constraints is not orthogonal, and again any
orthogonal basis is not localized and changes with time.

However, there is a (linear in the number of lattice points) 
algorithm for implementing
this update scheme approximately, which can be made highly accurate.  The
idea is that any update of form
\begin{equation}
E_i \rightarrow E_i - \frac{\mu \Delta t}{2 \pi^2} \left( B_i - \sum_\alpha
d_{\alpha i} \kappa_\alpha \right) \, ,
\end{equation}
with $d_{\alpha i}$ describing any basis for the Gauss constraints, and
with the $\kappa_\alpha $ completely arbitrary, 
will correctly modify the $E^*$.
It is not even essential that the basis for the Gauss constraints be 
orthogonal, as long as it is complete, independent, and orthogonal 
to all $E^*$.  What uniquely determines the right choices for the 
$\kappa_\alpha$ (once we have chosen a basis for the Gauss
constraints and hence the $d_{\alpha i}$)
is that the Gauss constraints must be satisfied at the end.  
If we can find an algorithm which, by only making changes to the $E$ fields
orthogonal to the $E^*$, incompletely but very accurately enforces the
Gauss constraints at each step, by choosing almost the correct
$\kappa_\alpha $, then we will have almost the evolution
of Eq. (\ref{rightway}); and by improving the accuracy to which the
Gauss constraints are approached we can test whether the results depend on
the residual failure.  This is the technique I will adopt for the
evolution of the Yang-Mills system with a chemical potential.  It is
still not clear whether the failure to correctly simulate the ultraviolet 
physics will influence the resulting $\Gamma_{\mu}$ (although
I will of course check if my results have sensible small volume behavior),
but the infrared evolution should be correct, and no spurious physics
should arise from violations of the Gauss constraints.

\begin{center}
 Restoring Gauss Constraints
\end{center}

\smallskip

I will now briefly describe two algorithms for the removal of the Gauss
constraint violations.  
For the uninterested reader it is only necessary to know
that algorithm 1 removes about half of the accumulated Gauss violation at
each step, and algorithm 2 is about twice as costly in compute time, but
leaves a residual Gauss constraint about one fifteenth as large.

The basic idea of quenching the Gauss constraint is to change the $E$ field
on a link
in the direction which will reduce the Gauss violation at both endpoints.
The most basic method is 
\begin{equation}
E_i(x) \rightarrow E_i(x) + \gamma (U_i C(x + \hat{i})U_i^{\dagger} - C(x) )
\end{equation}
where by $C(x)$ I mean the Gauss violation at the point $x$,
and I have of course parallel transported the forward $C$ to the point
where the group indicies of $E_i$ reside.  This algorithm is equivalent
to the relaxation algorithm with Hamiltonian $H = \sum_x C(x) \cdot C(x)$ (Lie
algebra dot product) used in \cite{Ambjornetal}, and it does not alter
the $E^*$.  Algorithm 1 is
to apply this relaxation once at each timestep, 
using a value of $\gamma$ chosen to make it
efficient.  To evaluate its efficiency it is useful to use Fourier
analysis, which should be approximately valid in the ultraviolet (where
most of the Gauss violation occurs).  When $C(x) = C(k) \sin(k \cdot x)$,
each update takes
\begin{equation}
C(k) \rightarrow 
\left( 1 - \gamma \sum_i 2 ( 1 - \cos k_i ) \right) C(k) \, .
\end{equation} 
The sum is recognizeable as $\omega ^2 (k)$ for the lattice dispersion
relation, and its maximum value is 12.  The sign of $C(k)$ oscillates
with each application of the algorithm if $\gamma > 1/\omega^2(k)$, and
if $\gamma > 1/6$ then the most ultraviolet mode becomes unstable.  For
$\gamma$ close to 1/6 the damping becomes inefficient for the most
ultraviolet modes.  However, the damping of 
infrared modes is in general quite
inefficient, and it is good to have $\gamma$ as large as practical.  Using
$\gamma = 0.1$ or $0.12$, I find the algorithm is quite efficient.  If
the algorithm is applied at each step, then the
total Gauss violation, measured by $\sqrt{\sum_x C(x) \cdot C(x) }$, after a
step is about as large as the Gauss violation generated by one step, starting
from no violation.  For an $18^3$ grid, a timestep of $0.06$ lattice spacings,
a chemical potential $\mu = 0.2$ in lattice units,
and a lattice inverse temperature $\beta_L = 6$, I find $\langle
C(x)_{rms} \rangle \sim 1.7 \times 10^{-4}$.

To determine how much of the accumulated $N_{CS}$ in the realtime evolution
of the lattice system was due to the Gauss
constraint violation, I performed a linear regression fit of the energy
of a lattice system to $N_{CS}$ when the chemical potential term was
turned on with $\mu = 0.2$ in lattice units, at $\beta_L = 5$ ($\mu = T$).
The slope should equal $\mu$ if no Gauss constraint violations are 
generated; it was low by $7.5 \% $, indicating that a nontrivial minority
of the measured change in $N_{CS}$ was due to the violation and subsequent
quenching of Gauss constraints.  When the algorithm was only applied every
16 steps, the slope was low by $37 \% $.  Clearly, then, quenching the
Gauss constraints effectively makes a great deal of difference, but algorithm
1 is insufficiently thorough.

Algorithm 2 is a modification and improvement of algorithm 1.  Because
the time stepsize is by necessity much 
shorter than any frequency in the problem,
the Gauss constraint violation generated at each time 
step will approximately equal
that of the previous time step; a good first correction is to repeat some
multiple $m < 1 $ of the total modification made in the previous time
step.  Then one application of the relaxation algorithm described above
is applied.  This combination leaves an rms Gauss violation
a factor of 4 smaller than algorithm 1.  
A further improvement is accomplished by applying
the relaxation step twice each time step, with different stepsize
constants $\gamma_1$ and $\gamma_2$.  All that is necessary for algorithm
stabiltity is that $1 > (1 - \gamma_1 \omega^2)
(1 - \gamma_2 \omega^2) > -1/3$ (the lower bound would be -1 for $m = 0$, 
but changes when $m \neq 0$.  The value I use is for $m=1$.).
By making the $\gamma$ well separated, with $\gamma_2 >1/6$, I make the
algorithm efficient over a wide frequency range.  I find that the parameter 
choice $m = 0.9$, $\gamma_1 = 5/48$, 
and $\gamma_2 = 5/24$ gives excellent performance, with the Gauss 
constraint violation after application about 1/15 as large as with algorithm 1.
Using this combination, the linear regression fit of energy versus $N_{CS}$
gives $\mu$ to better than $1 \%$ for every lattice spacing used in
this paper.  I conclude
that evolving the system using algorithm 2 is for all practical purposes
equivalent to implementing Eq. (\ref{rightway}). 
Incidentally I can also conclude that I implemented the chemical potential
with the right numerical coefficient.

If it should prove necessary,
it is straightforward to further improve algorithm 2; one repeats the 
relaxation step more than twice, with a different stepsize constant 
each time, chosen so that the function $\Pi (1-\gamma_i \omega^2)$ will be
as near zero in as wide a range as possible, and never outside the
stability bound.  For instance, I found that 3 appliacations with 
$\gamma = 7/72$, 7/48, and 7/24 leaves a residual violation about 2.5 times
smaller than algorithm 2.  Also, reducing the stepsize improves algorithm
performance; algorithm 1 improves linearly with $\Delta t$ and algorithm 2
(keeping $(1-m)/\Delta t$ fixed) improves quadratically.

\section{Some Numerical Results}

I implemented the Hamiltonian system with chemical potential described
above, and the thermalization algorithm of Krasnitz\cite{Krasnitz}.  
The time stepsize for the Hamiltonian evolution 
was chosen as follows:  in the ultraviolet (where most
of the energy of the system resides) the Fourier modes behave as weakly
coupled harmonic oscillators with frequency given by the lattice dispersion
relation $\omega^2 = 2 \sum (1 - \cos k_i)$.  Defining the energy of the
system at time $t$ to be 
\begin{equation}
{\rm Energy}(t) = \sum_{\Box} 1 - \frac{1}{2} {\rm Tr} U_\Box (t) 
+ \sum \frac{E^2(t+ \frac{\Delta t}{2})}{4}
+ \sum \frac{E^2(t-\frac{\Delta t}{2})}{4} \, ,
\end{equation}
the leapfrog algorithm should keep the
central value of energy stable for a harmonic oscillator, but
the amplitude of oscillations in the energy of a harmonic oscillator
should be $\omega^2 \Delta t^2/4$ times the energy.
To keep these fluctuations $ \leq 1 \%$ for the most ultraviolet
mode ($\omega^2 = 12$) it is necessary to choose $\Delta t \leq 0.06$.
I choose $\Delta t = 0.06$ in all simulations discussed in this
paper.  Integrating
over all the modes, the perturbative estimate of the variance in 
the energy of an $N^3$ lattice at
lattice inverse temperature $\beta_L$, in the limit of large $N$, is 
\begin{equation}
\sigma_{\rm Energy}^2 \simeq  (N_{\rm DOF})(\langle E^2_{\rm DOF} \rangle )
	\frac{\langle \omega^4 \rangle (\Delta t)^4}{32} = 
6 N^3 \frac{2}{\beta_L^2} \frac{213}{8} \frac{(\Delta t)^4}{32}
 \simeq 1.29 \times 10^{-4} \frac{N^3}{\beta_L^2}
\end{equation}
for my value of $\Delta t$.  The energy of the system without chemical
potential was in fact absolutely stable to overall drift, with variance
within $10 \%$ of the analytic estimate, for runs on large and small 
lattices.  I
also checked that, for initial conditions with only one excited mode, the
dispersion relations were correct.

To test the thermalization algorithm I measured the overall energy and
compared it to $6 N^3/\beta_L$, which is the free field estimate (there are
6 degrees of freedom per site); it agreed to a few percent, better at
large $\beta_L$ and worse at low $\beta_L$.  I also compared the values of
Wilson loops to the results of Krasnitz\cite{Krasnitz}, and found good
agreement.

A more substantive check was to measure the $N_{CS}$ diffusion rate in
the absence of a chemical potential for a thermalized system, 
comparing to the results of Ambjorn
and Krasnitz.  Because my computer resources were limited I could not
check their results at the large values of $\beta_L$ they used; instead I
measured $\Gamma_d$ at a range of smaller $\beta_L$, to find at what value
lattice coarseness effects arise.  I was also unable to demonstrate in
a convincing way that the ultraviolet, white noise fluctuations satisfy
their analytic estimate of $\delta N_{CS}^2 = 0.00684 N^3 / (\pi \beta_L)^2$;
instead I assumed this behavior and subtracted it off from the value of
$(N_{CS}(t) - N_{CS}(0))^2$; however I compare two values of $t$ to check
the validity of this procedure.  Expressing $\Gamma_d = \kappa_d / (\beta_L
\pi)^4$, (this value of $\kappa$ corresponds to the value defined in
the introduction),
I present my values for $\kappa_d$ in Table 1.  All error bars
are one $\sigma$ and statistical.  

It is clear from the table that values of $\beta_L \leq 5$ are contaminated
with some finite lattice spacing effect, but the results above this scale
are consistent with those of Ambjorn and Krasnitz.  

Next I implemented the chemical potential on the lattice, as described
in the previous section.  As a first check to see whether the technique
will be plagued with ultraviolet divergences, I thermalized the system so
that all $U$ and $E$ would fall in an abelian subspace of SU(2) and its
Lie algebra.  This is easy to do with the thermalization algorithm of
Krasnitz, because initial conditions which only contain excitation in one
Lie algebra direction never have the other Lie algebra directions excited
by the algorithm.  The evolution is then equivalent to compact U(1) gauge
theory, in which $N_{CS}$ should oscillate about 0 but never drift.  Indeed,
when I applied a chemical potential, the value of $N_{CS}$ did not
drift, but fluctuated in a narrow range about 0; 
so if there are ultraviolet problems in the chemical potential 
method, they only arise out of nonabelian interactions.

As a next test of the reliability of the
algorithm I measured $ \langle \dot{N}_{CS} \rangle $ 
with $\mu = 1/\beta_L$, which should
give half $\Gamma_d V$.  I will express my results as $\kappa(\mu \beta_L)
\equiv \langle \dot{N}_{CS} \rangle 
(\beta_L \pi)^4/(\mu \beta_L N^3)$; we should expect
$\kappa(\mu \beta_L) = \kappa_d / 2$ 
for small $\mu \beta_L$.  I find that, for
$\beta_L = 5$, $\kappa(1.0) = .46 \pm .03$; for $\beta_L = 6$,
$\kappa(1.2) = .50 \pm .04$; and for $\beta_L = 8$, $\kappa(1.0) = .60 \pm
.11$.  All three results are on lattices of length $N = 3 \beta_L$, 
well larger than the size Ambjorn and Krasnitz found would remove
finite scaling effects.  These results are in very good agreement with
the measured values of $\kappa_d$ at the same values of $\beta_L$,
which is suggestive that both techniques are working correctly.

However, since the implementation of the chemical potential, while 
probably correct for infrared excitations, is almost certainly wrong
for ultraviolet excitations at the lattice scale, it is a good idea to
test the finite volume behavior of the results.  We expect that, when the
volume of the lattice is made small, there will not be enough room for
energetically unsuppressed topology changing events, which must be
spatially extensive, to exist, and the 
rate of topology change should fall.  If our results arise from true
infrared behavior then we should see $\kappa(\mu \beta_L)$ fall; but
if our results arise from spurious
ultraviolet effects, then the rate should simply scale with the 4-volume
of the simulation and $\kappa(\mu \beta_L)$ should be unchanged.

\begin{figure}
\centerline{\psfig{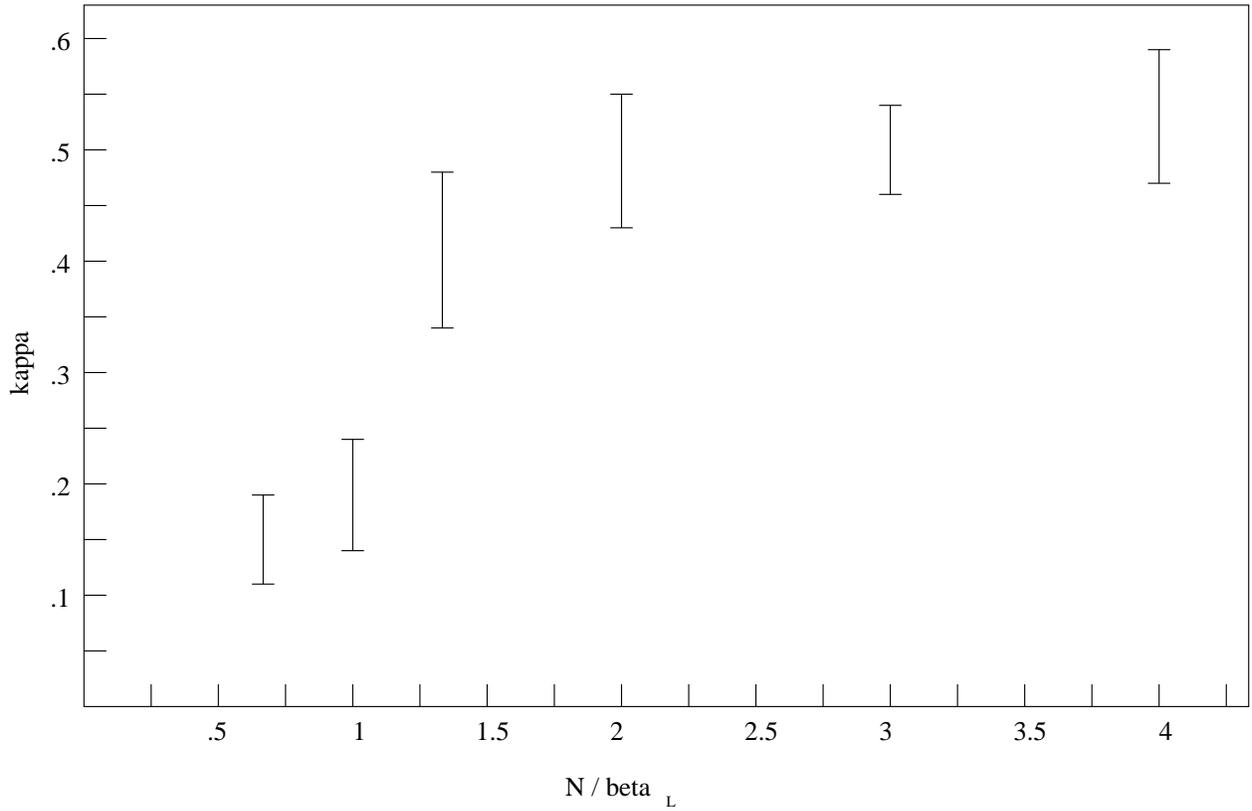}}
\caption{Lattice size dependence of the rate $\dot{N}_{CS}$.  The
horizontal scale is $N/\beta_L$, the vertical scale is 
$\kappa(\mu \beta_L)$.  If the
rate arose out of ultraviolet physics then the curve would be flat. 
\label{blah}}
\end{figure}

I measured $\kappa(\mu \beta_L)$ for $\mu \beta_L = 1.2$ and $\beta_L = 6$
for a number of lattice sizes.  The results are presented in Table 2 and
Figure \ref{blah}.  Each datapoint represents a total lattice 4-volume
of about $2 \times 10^{7} a^4$ ($a$ the lattice length).
It is clear that the rate is falling for small lattice
size, and the curve is about the same as that found by Ambjorn and Krasnitz
for $\kappa_d$.  Hence we can conclude that the results are in fact due
to infrared physics, which the model should treat properly.

\begin{figure}
\centerline{\psfig{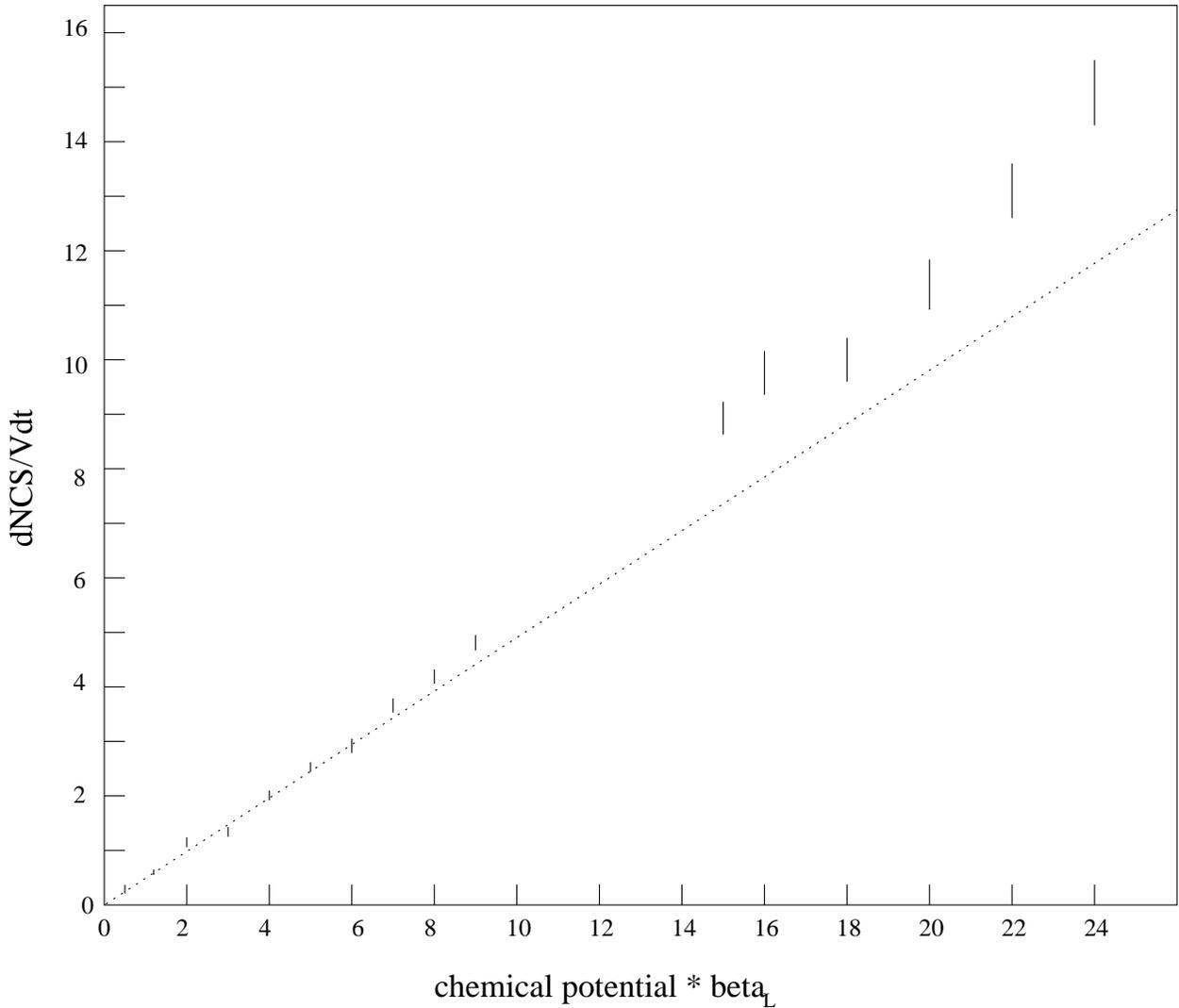}}
\caption{Dependence of $\dot{N}_{CS}$ on $\mu$.  Horizontal scale is
$\mu \beta_L$, vertical scale is $\dot{N}_{CS} (\pi \beta_L)^{4}/N^3$.
The line is a reasonable best fit to the points at lower $\mu$. 
The dependence is
startlingly linear up to $\mu \sim 6/\beta_L$, demonstrating that there
is not a free energy barrier to change in $N_{CS}$. \label{blah2}}
\end{figure}

With the algorithm thus tested, I can now investigate a new question; what
is the behavior of the system under large chemical potentials $\mu \gg
1/\beta_L$?  This directly tests whether there is a free energy barrier
to the flow of $N_{CS}$, as I discussed in section 2.  Again, I use
$\beta_L = 6$ in a compromise between compute time (which goes as
$\beta_L ^ 4$) and lattice refinement.  I present some results in
Table 3 and plot them in Figure \ref{blah2}.  For the larger values 
of $\mu$ it was necessary to correct for the finite heating of the 
system, and to use several short evolutions of independent thermal
initial conditions to minimize the amount of heating.
The data show that $\langle \dot{N}_{CS} \rangle $ rises almost linearly
with $\mu$ up to $\mu \beta_L \geq 6$, in stark contrast to the 
behavior when there is a free energy barrier.  The errors quoted are
all statistical, but there may be common systematic errors at the level
of $3 \%$ from thermalization and finite stepsize.
I also measured $\langle \dot{N}_{CS} \rangle$ 
for $\beta_L = 10$, $\mu \beta_L = 10$, and $N = 30$ and found
$\kappa(10) = .607 \pm .022$, which shows that the approximate linearity
is not a small $\beta_L$ artefact, but that the slight departure
from linearity (the excess of $\kappa(10)$
over $\kappa(1)$), is also probably real.

\section{Modified Lattice Hamiltonians}

There is one potential problem with the classical lattice technique which
might mean that the results for $\Gamma_d$ and $\Gamma_{\mu}$
do not correspond to the correct continuum, quantum behavior.  While
the thermodynamic properties of the lattice system may be the same as the
continuum theory in the infrared
(in the very good approximation of dimensional reduction
for the continuum theory), the dynamics may be different; in particular,
while we know that the ``hard thermal loop'' corrections to the static
propagator, namely the Debye screening, are of the same functional form
as in the quantum system, this does not extend to the nonzero frequency
case.  Bodeker
et. al. have recently shown that, because the most ultraviolet modes
move under lattice rather
than ultrarelativistic dispersion relations and have a rotationally
noninvariant ultraviolet cutoff, they induce
corrections in the propagators of the intermediate scale which are not
the same as the ``hard thermal
loop'' effects of the quantum theory, and are in fact not even rotationally
invariant \cite{McLerranetal}\footnote{In fact Bodeker et. al. show this
explicitly only for the abelian Higgs theory, but there is no reason to
doubt that the same thing happens in Yang-Mills theory as well.}.  
This sort of cutoff dependence is typical
of linearly divergent quantities.  While the lattice coarseness independence
of $\Gamma_d$ demonstrates that the overall magnitude of the ``hard
thermal loop'' effects does not matter (at least when the lattice is
fine enough that the Debye screening scale and the nonperturbative
scale are well separated), 
it is possible that the detailed $\vec{k}/k_0$ dependence of these corrections
matters in the determination of $\Gamma_d$.  This possibility is motivated
by Braaten and Pisarski's calculation of the gluon damping rate, 
in which the magnitude of the plasmon mass (proportional to the strength of
hard thermal loop effects) cancels out, but the functional form of
the hard thermal loops determines the calculation\cite{Braaten}.

A related but less urgent matter is to understand why $\Gamma_d$ falls
as the lattice becomes quite coarse.  There are two possible explanations for
this behavior.  One is that the lattice is incorrectly evolving the 
infrared equations of motion due to artefacts, namely nonrenormalizeable
operators induced by the crude choice of lattice action.  As the lattice
becomes finer the effect of these artefacts on the behavior of the 
nonperturbative infrared modes falls as $\beta_L^{-2}$.  Another possibility
is that $\Gamma_d$ is only constant in the limit that the nonperturbative
length scale ($O(\beta_L)$ in lattice units) and the Debye screening 
length scale ($O(\beta_L^{1/2})$ in lattice units) are well separated.

One can study at least the first of these issues by the following technique.
I design a lattice Hamiltonian which produces 
different (wrong) ultraviolet dispersion
relations and will induce different (wrong) hard thermal loops.  Then I 
measure $\kappa(\mu)$ for different lattice coarsenesses and see if there
is a large lattice size limit and whether it is the same as the limit
for the normal cubic lattice action.  If it is, then the limit did not
depend on which (wrong) hard thermal loops were induced, and would presumably
not change if the right hard thermal loops could be induced.

I examined two modifications of the lattice action.  In the first, I make
the spatial part of the Hamiltonian 
$5/3 \sum_\Box (1 - 1/2 {\rm Tr} U_{\Box})
- 1/12 \sum_{\sqsubset \! \sqsupset} (1 - 1/2 {\rm Tr} 
U_{\sqsubset \! \sqsupset }) $, where the second sum is over
all $1 \times 2$ rectangular plaquettes (both orientations).  This choice
gives steeper dispersion relations in the ultraviolet.  For modes with
$\partial_1 E_1 = \partial_2 E_2  = \partial_3 E_3 = 0$ the expansion
of $\omega^2$ in $k$ has no quartic term; but for other modes it has
positive $k_1^2 k_2^2$ type terms, and the action is not an ``improved''
action in the sense of eliminating dimension 6 operators from the 
Hamiltonian\footnote{The easiest way to see this is to consider the
thermodynamics of the model; introduce Lagrange multipliers $A_0(x)$ for
the constraint at site $x$ and perform the Gaussian integration over
the fields $E$.  This generates a kinetic term for $A_0$ which is
precisely the minimal (unimproved) implementation of $(D_i A_0)^2 /2$.
An improved action should produce an improved kinetic term for $A_0$.}.

I analyzed this model for several lattice coarsenesses (values of $\beta_L$),
generating the initial conditions with the thermalization algorithm 
discussed in Appendix A.  Because the Hamiltonian is more complicated,
the numerics were more time consuming; to get good statistics I was forced
to assume that $\langle \dot{N}_{CS} \rangle $ 
rises linearly with $\mu$ up to $\beta \mu \simeq 5$,
rather than verifying it explicitly as I did for the standard Hamiltonian.
Some results are presented in Table 4.  The same scaling behavior is 
reached, although the lattice must be finer before it is attained.  It is
not clear whether the approach to scaling is slower because the Debye mass
is smaller at the same $\beta_L$ (it depends on the integral of 
$\omega^{-2}$, and $\omega$ is larger in the ultraviolet in this model),
or because of the different dimension 6 artefacts.

An alternate and stronger check is to deliberately make cutoff scale
lattice behavior even worse, so the hard thermal loops will look even
less like their correct values.  If the small lattice size limit is
unaffected then there can be little doubt that the functional form
of the hard thermal loops really is not important.

To do this I make one lattice direction longer than the other two.  That
is, defining the electric field $E$ by 
$U(t + \Delta t) = \exp( i \Delta t \tau \cdot E) U(t)$,
I take 
as my dot product 
\begin{equation}
A \cdot B = \sum_{\rm sites} l ( A_1 B_1 + A_2 B_2 ) + l^{-1} A_3 B_3 \, ,
\end{equation}
as my Hamiltonian
\begin{equation}
H = \frac{ E \cdot E }{2} + \sum_{\Box} 
l ( 1 - \frac{1}{2} {\rm Tr} U_{\Box_{12}}) 
+ l^{-1} ( 1 - \frac{1}{2} {\rm Tr}
U_{\Box_{13}} + 1 - \frac{1}{2} {\rm Tr} U_{\Box_{23}} ) \, ,
\end{equation}
as my magnetic fields
\begin{eqnarray}
8B_1^a & = & l^{-1} \sum_{8 \Box} \frac{1}{2}{\rm Tr}(-i \tau^a) 
U_{\Box_{23}} \, ,
\nonumber \\
8B_2^a & = & l^{-1} \sum_{8 \Box} \frac{1}{2}{\rm Tr}(-i \tau^a) 
U_{\Box_{31}} \, ,
\nonumber \\
8B_3^a & = & l \sum_{8 \Box} \frac{1}{2}{\rm Tr}(-i \tau^a) 
U_{\Box_{12}} \, ,
\end{eqnarray}
as my definition of $D_j F_{ji}$
\begin{eqnarray}
(D_j F_{j1})_a & = & l^{-2} \sum_{\Box_{13}} \frac{1}{2} {\rm Tr} (-i \tau_a)
U_{\Box_{13}} + \sum_{\Box_{12}} \frac{1}{2} 
{\rm Tr} (-i \tau_a) U_{\Box_{12}}
\, \nonumber \\
(D_j F_{j2})_a & = & l^{-2} \sum_{\Box_{23}} \frac{1}{2} {\rm Tr} (-i \tau_a)
U_{\Box_{23}} + \sum_{\Box_{12}} \frac{1}{2} 
{\rm Tr} (-i \tau_a) U_{\Box_{12}}
\, \nonumber \\
(D_j F_{j3})_a & = & \sum_{4 \Box} \frac{1}{2} {\rm Tr} (-i \tau_a)
U_{\Box} \, ,
\end{eqnarray}
(where the orientations should be taken as appropriate), and 
as my $\Delta N_{CS}$
\begin{equation}
\Delta N_{CS} = \frac{ E \cdot B \Delta t}{2 \pi^2} \, .
\end{equation}
The update rules for $U$ and $E$ are, in terms of $B$ and $D_j F_{ji}$ 
defined above, unchanged.  The thermalization algorithm and the Gauss
law restoration are discussed in Appendix A.  The thermalization is applied
so that the total energy will be $6 l N^3 / \beta$ (6 the number of degrees
of freedom).

To determine how we should expect this modification to change the observed
rate of $N_{CS}$ violation, it is simplest to successively stretch each
lattice direction by the same factor of $l$.  The resulting model is the
same as the original cubic lattice model, except for factors of $l$.  The
energy in the magnetic field goes as 
$l^{-1}\sum_\Box 1 - 1/2 {\rm Tr} U_\Box$, 
but its magnitude grows as $l^3$; so the magnitude of 
$1-1/2{\rm Tr} U_\Box$
has grown as $l^4$.  Therefore the new model corresponds to the old model
with $\beta_L$ divided by $l^4$.  Also, because the expectation value of
$E$ only grows as $l$ and not $l^2$, we find that the time step has 
changed by a factor of $l^{-1}$.  The chemical potential, in lattice units,
is unchanged.  Therefore, the expected value of 
$\langle \dot{N}_{CS} \rangle$
should change as the $4*3-1=11$ power of $l$, and stretching only one
direction by $l$, using the definition of the system discussed above,
should change the rate by $l^{11/3}$.  To compensate, I shift my
value of $\beta_L$ by $l^{-4/3}$ and my value of $\Delta t$ by $l^{1/3}$;
I should then expect the rate to be $l$ independent, as long as $\beta_L$
is large enough to eliminate finite spacing artefacts and $N$ is large
enough to eliminate finite volume artefacts.  Because the grid is
rectangular, these are more stringent conditions than in the cubic lattice
case.

I have tested the behavior of this streched lattice for several values of
$l$, choosing $\beta_L$ (including the $l^{-4/3}$ correction) to be around
8.  Some results are presented in Table \ref{stretchtable}.  The data 
suggests a very weak dependence on $l$, with the rate rising slightly
for quite high values.  This suggests some weak lattice shape dependence,
presumably arising from a weak dependence on the hard thermal loops, 
but the effect is small and it is reasonable to believe based on this
evidence that the value for $\kappa$ determined from the cubic lattice
should be quite close to the value we would find if we could include
the right hard thermal loops.

\section{SU(3)}

The topological structure which exists in SU(2) (weak isospin) 
gauge theory also exists for SU(3) (color); there is a Chern-Simons
number, and at high temperatures, thermal excitations should be the dominant
means of topology change\cite{McLerranShap}.  
The rate of these changes, $\Gamma_{\rm strong}$, is
another important quantity to baryogenesis, since it may control the
rate at which fermions change chirality in the plasma\cite{Giudice}.
Parametrically, the $N_{CS}$ diffusion rate can be written as
$\Gamma_{d,\rm strong} = \kappa_{d,\rm strong} (\alpha_s T)^{4}$, with the
constant $\kappa_{d,\rm strong}$ another unknown.

The rate can be measured by a straightforward extension of the chemical
potential method used for SU(2).  Group and Lie algebra elements should
be chosen from SU(3) rather than SU(2), and $1 - 1/2 {\rm Tr}$ should
be replaced by $3/2 - 1/2 {\rm Re \: Tr}$; with these substitutions
everything carries over.  The lattice inverse temperature is now related
to the physical temperature as $\beta_L = 4/ (g_s^2 aT)$,
which will generate the correct power of $g_s$; the constant 
$\kappa_{d,\rm strong} = 2 \kappa_{\mu, \rm strong}$ 
can be gotten from simulations
in the same way as $\kappa_{\mu}$.  The thermalization algorithm,
presented in Appendix A. also carries over to SU(3) without serious
modification.

In practice, because the group SU(3) is larger than SU(2), and because
the anticommutators of the Lie algebra generators $\lambda_a$ are not 
multiples of the identity element, it is much more numerically costly
to simulate the group SU(3); I find that the time to update a lattice
of the same size is approximately 8 times larger for SU(3) than for 
SU(2).  Also, I find that the inverse lattice spacing $\beta_L$ 
must be larger before the results become $\beta_L$ independent.  Because
of these complications, I was unable to verify that  $\dot{N}_{CS}$
rises linearly with $\mu$.  Instead, I have assumed that this behavior
carries over from the SU(2) case.  I can then use quite a large value
for $\mu \beta_L$, which is necessary to get good statistics out
of relatively short runs.  
(The length of a run necessary to get fixed statistical
error scales as the square of $\mu$.)  Unfortunately, this introduces a
possible systematic error; it can be eliminated only with a great deal
more computer time.

I have also been unable to check explicitly that the rate displays similar
lattice volume behavior to the SU(2) case; instead I have assumed that
the rate becomes volume independent on lattices with $N > 2 \beta_L$.  Since
the magnetic screening length of SU(3) is presumably shorter than 
that for SU(2), this assumption is almost certainly justified.

I present some numerical results for $\kappa_{\mu,\rm strong}$ in Table 
\ref{SU3table}.  Because of the limitations discussed above, these
should be considered preliminary; the systematic error bars are
probably at least $10 \%$.  A preliminary estimate for the ratio
$\kappa_{\mu,\rm strong}/\kappa_{\mu}$ is $7 \pm 1$, but it is not clear from
the data whether the finest lattice used was large enough to reach the
fine lattice limit.

Note also that the energy is systematically higher than the leading order
perturbative estimate (the number of degrees of freedom divided by $\beta_L$).
This effect is real.  Since the electric fields appear quadratically in
the Hamiltonian, I can compute that the energy in $E$ fields should be
$8 N^3 / \beta_L$, and this is satisfied numerically.  The same energy
excess in the magnetic fields was observed in SU(2); for SU(3)
it is more than twice as large, presumably because there are more
ways for the fields to interact.

\section{Conclusion}

It is possible to include a chemical potential for $N_{CS}$ in lattice
simulations of the classical Yang-Mills 
field equations, and the rate of $N_{CS}$ motion
is related to that of $N_{CS}$ diffusion in the manner demanded by the
detailed balance argument.  Furthermore, the results show that there is
no free energy barrier to topology change in the symmetric phase, because
the response of $N_{CS}$ grows linearly with chemical potential over
a much wider range than is possible if the transitions occur by
hopping over a barrier.
I have also demonstrated that the classical, lattice
rate is weakly, if at all, dependent on
the details of the hard thermal loops induced by the high momentum modes.

To conclude, I will mention a possible interpretation of the results, and
some directions for future work.

The two striking things about the rate of $N_{CS}$ change under a chemical
potential are that, in terms of the volume necessary to remove finite
size effects, the rate seems very low, and that the rate remains linear
up to surprisingly large values of chemical potential $\mu$.  These two
things may be related.  Consider a square Wilson loop of length $x$ on
a side in the continuum limit of the classical theory.  When $x << \beta $
(by which I mean $a \beta_L$ as we take $\beta_L$ to infinity and
$a$ to zero) the trace of the loop is $2(1-C x/(2 \beta)) $, where
\begin{equation}
C = \frac{n^2 - 1}{2 \pi^2} \int dk_1 dk_2 \frac{1-\cos(k_1)}{k_1^2}
\frac{1-\cos(k_2)}{k_2^2} \sqrt{k_1^2 + k_2^2} 
\end{equation}
for SU($n$) gauge theory.
This expression has an ultraviolet logarithmic divergence, but what I
am interested in is the contributions from $k \sim \pi /a$, which are
$O(1)$, with no powers of $\pi$.  The length scale where the ``correction''
term which makes the trace $\neq 2$ becomes order 1, which is the
scale where nonperturbative physics sets in, is $x \sim \beta_L$, with no
powers of $\pi$.  This is the length scale where excitations no longer
remain coherent long enough for energy to oscillate between $E$ and $B$
fields, and it is reasonable to expect, then, that it is the scale where
$\dot{N}_{CS}$ will lose its oscillatory behavior and behave in a Brownian
fashion.  The characteristic 4-volume of Brownian motion is thus 
$O( \beta^4)$, with no powers of $\pi$, which corresponds to the minimum 
volume for finite volume effects to disappear, and the
magnitude of $E \cdot B$, integrated over one such 4-volume, is $O(1)$,
again with no powers of $\pi$.  To get $\delta N_{CS}= E \cdot B /(2\pi^2)$
to be $O(1)$ then requires $O(\pi^4)$ characteristic Brownian motion 
4-volumes, and we should anticipate $\Gamma \sim (\pi \beta)^{-4}$,
not $\beta^{-4}$, or $\kappa_d \sim 1$, not $\sim \pi^4$.  

Turning to the
chemical potential, one would expect that $\mu$ would start seriously
changing the dynamics of the infrared when $\mu E \cdot B/(2 \pi^2)$, 
integrated over a volume of order $\beta^4$, is order $\beta^{-1}$, that
is, when $\mu \beta \sim \pi^2$; and that for values smaller than this,
the chemical potential is a small correction to the dynamics and the
response should be linear.  Again, this corresponds well to the actual
behavior of the system.

What possible future projects can the chemical potential method be 
applied to?  One question I could not answer here was whether
the $N_{CS}$ diffusion rate falls off on coarse lattices because the 
Debye screening length comes on order the scale of nonperturbative infrared
physics, or because of the effects of nonrenormalizeable
operators arising from the crudeness of the lattice action.  One
could resolve this question by finding a lattice Hamiltonian system
and an update algorithm which do not generate dimension 6 nonrenormalizeable
operators in the infrared physics.  If dimension 6 nonrenormalizeable
operators are causing problems on coarse lattices then such a program 
could also allow better study of the SU(3) case without requiring the
very fine lattices which make such study numerically expensive.
Also, knowing whether or not Debye screening is important would help
to clarify the role which hard thermal loops play in the calculation.

Also, extending the ``chemical potential method'' to the Yang-Mills Higgs
theory allows one to study the out of equilibrium decay of the infrared
gauge fields as they are swept up a bubble wall during the electroweak
phase transition.  It is frequently assumed in the literature that
any left handed baryon number asymmetry generated 
by the motion of the bubble wall must find its
way to the symmetric phase to cause baryon number generation, because
$N_{CS}$ violating processes are exponentially suppressed in 
almost all of the wall and
in the broken phase\cite{DineThomas}; and a contrary assumption, that
a chemical potential for $N_{CS}$ present on the wall can act on
the infrared gauge field configurations as they decay\cite{Neiletc}, has
never been convincingly tested.  The method proposed here could be
used to do so, for instance by integrating out the fermions analytically
to produce effective interactions in the two doublet model, and numerically
evolving the system through the electroweak phase transition.

\begin{center}
Acknowledgements
\end{center} 

I would like to thank A. Krasnitz for sparking my interest in this
topic, and for useful conversations and criticism, 
and Neil Turok and Chris Barnes,
for stimulating discussions.  I would also like to thank Carl Edman, for
donating a great deal of computer time.

\section{Appendix A}

In this appendix I present a simple, efficient thermalization algorithm
for the classical lattice Yang-Mills or 
Yang-Mills Higgs system, or for any system in which the
phase space is a vector bundle over the space of coordinates,
with a Hamiltonian which is a pure coordinate term
plus an inner product on the vector space, with the inner product allowed
to be coordinate dependent.  Constraints on the momenta are permitted
provided they are linear, ie in each vector space they restrict to a
vector subspace.
(In the case of Yang-Mills theory the phase space is the tangent bundle
over the space of $U$, and the constraints restrict the momentum vector
to have a zero projection in each direction which corresponds
to a gauge transformation
on the space of $U$.)
The algorithm is faster than that of
Krasnitz, and much easier to apply to new systems, although the above
requirements make it less general.
Almost all the numerical equipment it uses must be developed
anyway for the evolution of the system with a chemical potential.

The idea is the following; we begin in the vacuum.  Then we fully
thermalize the $E$ fields, leaving the link matricies $U$ untouched.  
That is, we draw $E$ with correct Boltzmann weight from the vector space of 
allowed $E$ at that value of $U$.  We then
evolve the fields according to the classical Hamiltonian equations of
motion, allowing the thermalization to mix between $E$ and $U$ fields.
Then we throw out the values of the $E$ fields and fully rethermalize
them; we repeat the thermalization and evolution over and over until the
system is completely thermalized.  

To see that this is a correct thermalization algorithm,
consider the evolution under this algorithm, averaging over realizations,
of an ensamble of systems described by some probability
distribution.  If we really draw $E$
with correct thermal weight from the fixed $U$ fiber (the set of $E$ at
fixed $U$),
then the correct thermal probability distribution will be preserved by
the algorithm.  Further, any probability distribution preserved by
the algorithm must have the property that the probability distribution
is a product of a function on the coordinate manifold and a function
on each fiber, with the function on each fiber the thermal
distribution for the $E$ fields at that value of $U$; 
and this property must be
preserved by the Hamiltonian evolution.  Only the correct
thermal distribution satisfies these conditions.  Hence the algorithm 
should be correct.  The only problem is finding a way to draw $E$ from
the thermal distribution on a fiber.

Now the thermal distribution function on the
fiber of a particular $U$ (that is, on the space of $E$ at fixed $U$) is
Gaussian.  All we have to do to thermalize the $E$ fields in this
vector space is draw a value from the Gaussian distribution.
In the absence of constraints, this would be easy.  
The problem is that, in the obvious natural basis for the $E$ fields,
the constraints are nontrivial linear combinations of $E$ fields,
and to find an orthonormal basis of the directions in the dynamical
subspace (the subspace permitted under the constraints) we must diagonalize
a large matrix, as discussed in section 3.  However, this does
not turn out to be a fundamental problem.

The key is to notice that the obvious basis of $E$ fields is orthonormal
(up to a factor of 2) under the inner product defined by the Hamiltonian.
If we choose each $E$ field independently from the Gaussian distribution,
then that will also choose the fields independently from the Gaussian
distribution in any other orthonormal basis for the $E$ fields, a special
property of the Gaussian distribution.  This includes
any basis which decomposes into  a subset 
$\{ E^* \}$ orthogonal to the constraints
and a subset $\{E^c \} $ along constraint directions.
This means that we
will have chosen the $E^*$ correctly.  All that remains is to set the
$E^c$ to zero without changing the $E^*$, that is, to orthogonally
project to the constraint subspace.  But I have already
presented a dissipative algorithm for doing so; by applying the
dissipation step repeatedly, we can quite easily bring the magnitudes 
of the Gauss constraint violations to the level of machine roundoff
accuracy.  The electric fields are then correctly thermalized.

I should comment that the demands on the dissipative algorithm here
are different from the demands during the evolution of the fields with
chemical potential; so the optimal values of the algorithm coefficients
are quite different.  I find that, for algorithm 2, the value of
$m$ should be brought down to $m \leq 0.5$.

I implemented this algorithm for the Yang-Mills system and compared it
to the algorithm of Krasnitz\cite{Krasnitz}; 
I used the stepsize $\Delta t = 0.03$ in
lattice units for my algorithm and $\Delta t = \gamma_E = 0.03$ for
the algorithm of Krasnitz.  I used 10 $E$ field randomizations with evolutions
of length $\beta_L$ between them for my algorithm.  On lattices of $N=15$
and $N=18$ I systematically found my algorithm gave total energies about
$1 \%$ less than the algorithm of Krasnitz; this is probably because of
the rather large time step and dissipation constant I used to implement
his algorithm.  (In his paper he advocates a more conservative stepsize.)
I also compared the values of Wilson loops and found them
to be within the same $1 \%$ margin.  That their values are preserved
by the Hamiltonian evolution of the thermalized system, is more or less
automatic in my algorithm; but I tested it anyway and found it was true.

It is completely straightforward to extend this algorithm to the group
SU(3); my results for SU(3) use my thermalization algorithm.  It
is also reasonably straightforward to apply it to SU(2) Higgs theory; but
here it must be remembered that the constraint is a linear combination
both of $E$ fields and of $\pi$ fields (the conjugate momentum of the
Higgs doublet); the dissipation algorithm used in the thermalization 
should change both fields in
proportion to their contribution to the Gauss constriant (holding $U$
and $\phi$ fixed throughout the enforcement of the constraint).  The same
algorithm must be used in the evolution of the system under a chemical
potential.

It is never necessary
to dissipatively enforce the Gauss constraint during the evolution step
of the thermalization algorithm, because no chemical potential is applied
and the equations of motion preserve the constraint identically, as I
have of course verified numerically, for cubic and distorted lattices
and for SU(2) and SU(3).

When applying the algorithm to the distorted lattice, it
must be remembered to define orthogonality in terms of the quadratic
part of the Hamiltonian.  In other words,
since Energy=$l ( E_1^2 + E_2^2 + ( l^{-1}E_3)^2)$, when we find the
violation of the Gauss constraint $C_x = E_1 + E_2 + l^{-1} ( l^{-1} E_3)
-$(incoming lines), we should apply a correction of $- \gamma C_x$ to 
both $E_1$, and
$E_2$, and of $-\gamma (l^{-1} C_x)$ to $(l^{-1}E_3)$; so although
$l^{-2}E_3$ enters the Gauss constraint, $E_3$ is modified by 
$-\gamma C_x$, with
no power of $l$.  The same dissipation algorithm should be used during 
the evolution of the fields with chemical potential as in the dissipative
part of the thermalization, because the inner product defined by the
dot product is the same as that defined by the Hamiltonian.  

As a check that the algorithm correctly chooses $E$ fields for the 
elongated lattice system, I thermalized the
system with $l = 1.5$, ending by randomizing the $E$ fields.  I then allowed
the system to evolve under the Hamiltonian equations of motion discussed
in section 5, and checked to see if the distribution of energy between
$E_1 , E_2$ and $E_3$ changed; it did not, to within statistical errors of
less than $1 \% $, showing that the thermalization
correctly partitioned energy between these fields.  (If I shift $E_3$ by
$- l^{-2} C_x$ in the dissipation algorithm, which would be appropriate 
if I used the obvious rather than the correct metric on the $E$ fields, 
then the energy distribution does shift.)  The relation between
Wilson loops in the normal and elongated directions also was not affected
by randomizing the $E$ fields and allowing the system to evolve.

I should comment that the thermalization problem cannot be solved
by the following, very simple (but wrong), algorithm.  
Since the plaquette part of
the Hamiltonian does not depend on the $E$ fields, and the $E$ field part
is quadratic but with constraints, one uses a conventional lattice gauge
theory algorithm to thermalize the plaquette part and then the above,
simple algorithm to thermalize the $E$ fields.  The problem is that the
constraints depend on the $U$ fields; if we integrate over the (Gaussian)
$E$ fields we get a residual which is $U$ field dependent and shifts the 
relative weights of different $U$ field configurations; in fact it is 
exactly the $A_0$ fields of the dimensionally reduced theory.  What we
could do is use a standard lattice thermalization algorithm to thermalize
the 3 dimensional system with plaquette action and $A_0$ fields with
zero (bare) mass, use the resulting $U$ fields, throw away the $A_0$
fields, and choose the $E$ fields as discussed above.
This also ilustrates the relation between the thermalization
algorithm presented here and the ``molecular dynamics'' Monte-Carlo
algorithm of lattice gauge theory.

\pagebreak

\begin{figure}
\centerline{\psfig{file=counterexample.epsi,width=\hsize}}
\caption{The configuration of non-identity links which will make the discrete
spacetime integral of $F \tilde{F}$ nonzero; $U1=1+i\epsilon \tau_1$, 
$U2 = 1 + i \epsilon \tau_2$, and $U3 = 1 + i \epsilon \tau_3$.  Going
around the plaquette containing $U1$ and $U2$ picks up an $O(\epsilon^2)$
contribution in the $\tau_3$ Lie algebra direction, which gives a nonzero
contribution when traced against the plaquette which goes backwards in
time, even though all links at large distances are the identity.
\label{counterexample}
}
\end{figure}

\begin{table}
\begin{tabular}{|c|c|c|c|c|c|c|} \hline
 $\beta_L$  & 3 & 4 & 5 & 6 & 7 & 8 \\ \hline 
$\kappa_{d,t = \beta_L \pi} $ & $.48 \pm .04$ & $.71 \pm .06$ & 
$.93 \pm .09$ & $ 1.16 \pm .10$ & $1.07 \pm .10$ & $1.14 \pm .12$ \\ \hline
$\kappa_{d,t = 2 \beta_L \pi}$ & $.49 \pm .06$ & $.67 \pm .08$ & 
$.93 \pm .13$ & $1.08 \pm .13$ & $1.04 \pm .14$ & $1.30 \pm .20$ \\ \hline
\end{tabular}
\caption{Values of $\kappa_d$, the dimensionless $\Gamma_d$, for various
inverse lattice spacings.  All datapoints are for lattices of size
$N = 3 \beta_L$ to prevent finite volume effects.
Finite spacing effects become important around
$\beta_L = 5$.}
\end{table}

\begin{table}
\begin{tabular}{|c|c|c|} \hline
$N$ & $N/\beta_L$ & $\kappa( 1.2 ) $ \\ \hline \hline
4 & .67 & .15 $ \pm $ .04 \\ \hline
6 & 1.0 & .19 $ \pm $ .05 \\ \hline
8 & 1.33 & .41 $ \pm $ .07 \\ \hline
12 & 2.0 & .49 $ \pm $ .06 \\ \hline
18 & 3.0 & .50 $ \pm $ .04 \\ \hline
24 & 4.0 & .53 $ \pm $ .06 \\ \hline
\end{tabular}
\caption{Dependence of $\kappa(\mu \beta_L)$ on lattice volume for $\beta_L
= 6$.  The rate of topology change is clearly suppressed on small volumes
and saturates around $N = 2 \beta_L$.  Error bars are statistical and
$1 \sigma$.}
\end{table}

\begin{table}
\begin{tabular}{|c|c|c|c|} \hline
 $N$ & $\mu \beta_L$ & $ \langle \dot{N}_{CS} \rangle 
(\beta_L \pi)^4/ N^3 $ & $\kappa ( \mu \beta_L)$ \\ \hline \hline
18  &  0.5  &  .29 $ \pm $ .08  &  .58 $ \pm $ .16 \\ \hline
18  &  1.2  &  .60 $ \pm $ .05  &  .50 $ \pm $ .04 \\ \hline
18  &  2    &  1.15 $ \pm $ .09 &  .57 $ \pm $ .05 \\ \hline
18  &  3    &  1.34 $ \pm $ .09 &  .45 $ \pm $ .03 \\ \hline
18  &  4    &  2.01 $ \pm $ .09 &  .50 $ \pm $ .02 \\ \hline
18  &  5    &  2.53 $ \pm $ .09 &  .51 $ \pm $ .02 \\ \hline
18  &  6    &  2.92 $ \pm $ .13 &  .49 $ \pm $ .02 \\ \hline
18  &  7    &  3.66 $ \pm $ .13 &  .52 $ \pm $ .02 \\ \hline
18  &  8    &  4.19 $ \pm $ .13 &  .52 $ \pm $ .02 \\ \hline
18  &  9    &  4.81 $ \pm $ .14 &  .54 $ \pm $ .02 \\ \hline
18  &  15   &  8.93 $ \pm $ .30 &  .60 $ \pm $ .02 \\ \hline 
18  &  16   &  9.76 $ \pm $ .40 &  .61 $ \pm $ .02 \\ \hline
18  &  18   &  10.0 $ \pm $ .4  &  .56 $ \pm $ .02 \\ \hline
18  &  20   &  11.38 $ \pm $ .46 & .57 $ \pm $ .02 \\ \hline
18  &  22   &  13.1 $ \pm $ .5 &   .60 $ \pm $ .02 \\ \hline
18  &  24   &  14.9 $ \pm $ .6 &   .62 $ \pm $ .02 \\ \hline
\end{tabular}
\caption{Rate of $N_{CS}$ change as a function of chemical potential.
The rate is surprisingly linear, but turns up somewhat for $\mu \beta_L
\geq 10$.  Error bars are statistical and $1 \sigma $.}
\end{table}

\begin{table}
\begin{tabular}{|c|c|c|c|c|} \hline
Hamiltonian  &  $\beta_L$ & $N$ & $\mu \beta_L$ & $\kappa$ \\ \hline \hline
new  &  3.0  &  9  &  3.0  &  $.177 \pm .011$ \\ \hline
new  &  4.0  & 12  &  3.0  &  $.281 \pm .021$ \\ \hline
new  &  5.0  & 15  &  3.0  &  $.302 \pm .032$ \\ \hline
new  &  6.0  & 18  &  3.0  &  $.357 \pm .030$ \\ \hline
new  &  8.0  & 24  &  5.0  &  $.393 \pm .034$ \\ \hline
new  & 10.0  & 30  & 10.0  &  $.585 \pm .022$ \\ \hline
old  & 10.0  & 30  & 10.0  &  $.607 \pm .022$ \\ \hline
\end{tabular}
\caption{$N_{CS}$ violation rate for Hamiltonian including rectangular
plaquettes.  All volumes are large enough to eliminate finite volume
effects.  The rate begins to scale with lattice coarseness more slowly
reaches the same limit, as shown in the last column.}
\end{table}

\begin{table}
\label{stretchtable}
\begin{tabular}{|c|c|c|c|c|} \hline
$\beta_l$  &  $N$  &  $l$  &  $\beta \mu$  &  $\kappa({\mu})$  \\ 
\hline \hline
5.45  &  24  &  1.33  &  4  & .56$ \pm$.03  \\ \hline
6.70  &  24  &  1.14  &  4  & .50$ \pm$.04  \\ \hline
8.00  &  24  &  1.50  &  6  & .73$ \pm$.05  \\ \hline
8.00  &  24  &  1.50  &  3  & .67$ \pm$.05  \\ \hline
8.00  &  24  &  1.40  &  6  & .64$ \pm$.04  \\ \hline
8.00  &  24  &  1.30  &  6  & .64$ \pm$.04  \\ \hline
8.00  &  24  &  1.00  &  4  & .57$ \pm$.05  \\ \hline
8.00  &  24  &  0.80  &  6  & .57$ \pm$.04  \\ \hline
8.00  &  24  &  0.70  &  6  & .54$ \pm$.04  \\ \hline
8.00  &  24  &  0.60  &  6  & .55$ \pm$.04  \\ \hline
9.36  &  24  &  0.88  &  4  & .65$ \pm$.07  \\ \hline
10.77 &  24  &  0.80  &  4  & .47$ \pm$.08  \\ \hline
12.23 &  24  &  0.73  &  4  & .60$ \pm$.10  \\ \hline
\end{tabular}
\caption{An assortment of data on an elongated lattice.  There is no strong
trend, although the very large $l$ points turn up.}
\end{table}

\begin{table}
\label{SU3table}
\begin{tabular}{|c|c|c|c|c|} \hline
 $\beta_L$ & $N$ & Energy/$(16N^3/\beta_L)$ & $\mu \beta_L$ &
 $\kappa_{\rm strong}(\mu)$ \\ \hline \hline
  3  &  9   &  1.09   &  3  &  0.52$\pm$.05  \\ \hline
  4  &  12  &  1.10   &  4  &  1.19$\pm$.06  \\ \hline
  5  &  15  &  1.09   &  5  &  1.87$\pm$.09  \\ \hline
  6  &  18  &  1.07   &  6  &  2.30$\pm$.11  \\ \hline
  7  &  18  &  1.05   &  5  &  3.01$\pm$.19  \\ \hline
  10 &  24  &  1.035  &  6  &  3.19$\pm$.25  \\ \hline
  12 &  24  &  1.032  &  8  &  3.86$\pm$.30  \\ \hline
  14 &  28  &  1.025  &  6  &  4.26$\pm$.45  \\ \hline
\end{tabular}
\caption{Rate of topology change in SU(3) lattice gauge theory at 
various lattice refinements.  It is not clear that the fine lattice
limit is reached, even at twice the refinement necessary in the $SU(2)$
case.}
\end{table}

\end{document}